\documentclass{aastex6}

\usepackage[utf8]{inputenc}
\usepackage{graphicx}
\usepackage{bm}
\usepackage{color}      
\usepackage{units}
\usepackage{amsmath}

\begin{document}        
   
\title{\bf Particle-in-cell and weak turbulence simulations of plasma emission}

\author{Sang-Yun Lee}
\email{lizagd@khu.ac.kr}
\affiliation{School of Space Research, Kyung Hee University, Yongin, Gyeonggi 17104, Korea}

\author{L. F. Ziebell}
\email{luiz.ziebell@ufrgs.br}
\affiliation{Instituto de F\'isica, UFRGS, 91501-970 Porto Alegre, RS, Brazil}

\author{P. H. Yoon}
\email{yoonp@umd.edu}
\affiliation{School of Space Research, Kyung Hee University,
Yongin, Gyeonggi 17104, Korea} 
\affiliation{Institute for Physical Science and Technology,
University of Maryland, College Park, MD 20742-2431, USA}
\affiliation{Korea Astronomy and Space Science Institute,
Daejeon 34055, Korea} 

\author{R. Gaelzer}
\email{rudi.gaelzer@ufrgs.br}
\affiliation{Instituto de F\'isica, UFRGS,
91501-970 Porto Alegre, RS, Brazil}

\author{E. S. Lee}
\email{eslee@khu.ac.kr}
\affiliation{School of Space Research, Kyung Hee University,
Yongin, Gyeonggi 17104, Korea}

\date{\today}

\begin{abstract}
The plasma emission process, which is the mechanism for solar type II and type III radio bursts phenomena, is studied by means of particle-in-cell and weak turbulence simulation methods. By plasma emission, it is meant as a loose description of a series of processes, starting from the solar flare associated electron beam exciting Langmuir and ion-acoustic turbulence, and subsequent partial conversion of beam energy into the radiation energy by nonlinear processes. Particle-in-cell (PIC) simulation is rigorous but the method is computationally intense, and it is difficult to diagnose the results. Numerical solution of equations of weak turbulence (WT) theory, termed WT simulation, on the other hand, is efficient and naturally lends itself to diagnostics since various terms in the equation can be turned on or off. Nevertheless, WT theory is based upon a number of assumptions. It is, therefore, desirable to compare the two methods, which is carried out for the first time in the present paper with numerical solutions of the complete set of equations of the WT theory and with two-dimensional electromagnetic PIC simulation. Upon making quantitative comparisons it is found that WT theory is largely valid, although some discrepancies are also found. The present study also indicates that it requires large computational resources in order to accurately simulate the radiation emission processes, especially for low electron beam speeds. Findings from the present paper thus imply that both methods may be useful for the study of solar radio emissions as they are complementary.
\end{abstract}
    
 

\section{Introduction}

Solar radio bursts phenomena in the meter wavelengths were discovered and subsequently classified into five categories, starting from the 1950s and are still being studied to this date \citep{WildMcCready50/09, Wild50/09, Wild50/12, Wild+54/09}\@. The present work relates particularly to types II and III radio bursts. Both phenomena are related to high-energy beams of electrons produced during solar flares and Coronal Mass Ejections, which occur in the solar chromosphere and corona \citep{Benz17/12}, and are characterized by emissions in the VHF band, with frequencies ranging from $\unit[10]{MHz}$ to approximately $\unit[300]{MHz}$ (hence with wavelengths of the order of meters)\@. Dynamic spectra of both types II and III typically display negative frequency drift with time, but while the type III typically occurs in short bursts lasting a few minutes, the type II is characterized by a much slower downward drift rate, which can last for several tens of minutes \citep{Ergun+98/08, Bastian+98/09, Cane+02/10, Reiner+09/10, Hudson11/01, ReidRatcliffe14/07}.

The generally accepted mechanism by which types II and III solar radio bursts are generated was initially proposed by the pioneering work of \citet{GinzburgZheleznyakov58/10} and has been greatly improved by several contributions since then. According to the standard theory, disruptive events occurring in the solar chromosphere, such as solar flares, create bursts of energetic electrons that stream outwards from the Sun, along open magnetic fields lines. As these bursts propagate through the corona, which already contains a denser population of particles in a quasi-thermalized state, a beam-plasma instability is triggered when the speed of the beam electrons exceeds the thermal velocity of the background electron population. This bump-on-tail instability then excites the exponential growth of dispersive Langmuir waves ($L$ mode waves), starting from the initially low level of local thermal radiation field, which are predominantly emitted in the same direction as the beam. As the Langmuir waves are convectively amplified along their ray paths, the wave intensity becomes sufficiently high for nonlinear processes to start taking place. According to the weak turbulence theory of plasmas, the most important nonlinear processes in this phase are the three-wave decay/coalescence and nonlinear wave-particle scattering, which will first amplify the backscattered Langmuir mode from thermal background, mostly due to the coupling of the forward-propagating $L$ mode with the ion-acoustic ($S$) mode. At later stages of the nonlinear turbulent processes, when the backward $L$ mode reaches a sufficiently high intensity level, subsequent wave-wave and wave-particle interactions promote its coupling with the forward $L$ mode, thereby exciting the electromagnetic (transverse, $T$) mode. Longer time-scale processes can finally lead to the excitation of harmonics of the transverse mode. The entire series of events, which was broadly described above, is known in the literature as the plasma emission process and is supported by several contributions published in the literature \citep{Tsytovich67, KaplanTsytovich68/06, Melrose82/04, GoldmanDuBois82/06, Goldman83/12, Melrose87/03, Cairns87/10a, RobinsonCairns98/08a, RobinsonCairns98/08b, Zlotnik+98/03, Kontar01/04, Kontar01/08b, Kontar01/08c, Gosling+03/07, Li+05/01, Li+05/05, Li+06/09, Li+08/06, Li+11/03, Li+12/07, Ratcliffe+12/12, ReidKontar17/02}.

The plasma emission mechanism has also been studied within the framework of the Generalized Weak Turbulence (GWT) theory \citep{Yoon00/12, Yoon05/04, Yoon06/02, Yoon+12/10}\@. In this theory, the evolution of the turbulence level in a thermal plasma is described by a set of kinetic equations for the velocity distribution functions of the plasma species and for the spectral wave intensities of the various normal modes interacting with the particles. Examples of applications of the GWT theory to the study of the nonlinear evolution of the beam-plasma instability, taking into account wave-wave and nonlinear wave-particle interactions, are given by \citet{ZiebellGaelzerYoon01/09, ZiebellGaelzerYoon08/03, Gaelzer+08/04, Ziebell+08/08, Ziebell+11/01, Ziebell+11/08, Ziebell+12/05, Yoon+12/08, Ziebell+14/01a, Ziebell+14/01b, Ziebell+14/11, Ziebell+15/06, Ziebell+16/02}\@. In particular, in \citet{Ziebell+14/11} and \citet{Ziebell+15/06}, all the steps involved in the plasma emission process are considered. In addition to the wave kinetic equations for the forward/backward $L$ and $S$ modes, containing all the usual linear and nonlinear interactions included in the traditional weak turbulence theory, the above-mentioned publications also included the kinetic equations for the forward/backward transverse modes. The full set of kinetic equations for particles and waves is self-consistently solved during a sufficiently long evolution time that shows not only the excitation of the fundamental $T$ mode, but the first two transverse harmonic modes as well. The results obtained by \citet{Ziebell+14/11} and \citet{Ziebell+15/06} give additional support to the hypothesis that the underlying generating mechanism of the Type II/III solar radio bursts is the plasma emission process.

In another theoretical front, some authors applied the technique of particle-in-cell (PIC) simulation to the question of the plasma emission as the mechanism behind solar radio emissions. Some examples of contributions in this front are given by \citet{Kasaba+01/09}, \citet{KarlickyVandas07/12}, \citet{Rhee+09/03, Rhee+09/01}, \citet{Ganse+12/06,Ganse+12/10}, and \citet{ThurgoodTsiklauri15/12}\@. 

Until now, the few attempts that have been made for a direct comparison, either qualitative or quantitative, between the results obtained from the weak turbulence theory and from numerical simulations have all been of a limited scope. \citet{Rha+13/09} employed PIC simulation to test a theory for the generation of asymmetric superthermal tails in the electronic distribution function. According to the theory \citep{Yoon+12/08}, asymmetric tails can be created due to the nonlinear interactions of electrons with $L$ and $S$ waves, when the ion-electron temperature ratio $\left(T_{i}/T_{e}\right)$ varies within the range $0.1\lesssim T_{i}/T_{e}\lesssim1$\@. \citet{Ziebell+14/01a} employed simulation to test a theory for the generation of a quasi-thermal electromagnetic radiation field, resulting as the time-asymptotic state of the nonlinear interaction of the transverse mode with the longitudinal modes and particles, in the absence of free energy sources (\emph{e.g.}, beams) and collisions (\emph{i.e.}, no Brehmsstrahlung emission)\@. On the other hand, \citet{Ratcliffe+14/12} performed a direct comparison of the plasma emission mechanism with a PIC simulation. However, the weak turbulence formulation that was tested was one-dimensional and the Langmuir wave kinetic equation contained only the three-wave decay term, besides the usual quasi-linear diffusion term. Moreover, the transverse mode kinetic equation was absent in their comparison.

Here, we report for the first time a detailed comparison between the full two-dimensional (2D) plasma emission theory with a 2D particle-in-cell (PIC) simulation. The complete numerical solutions of 2D equations of generalized weak turbulence (GWT) theory can be termed the weak turbulence (WT) simulation, which are directly compared against numerical results obtained from the 2D PIC code simulation. Employing the same sets of physical parameters, the plasma emission process is simulated by both the 2D relativistic electromagnetic (EM) PIC simulation, employing a physical proton-to-electron mass ratio, and the GWT theory, employing the full set of 2D self-consistent kinetic equations for the particles and for $L$, $S$ and $T$ waves. The results show a reasonably good quantitative agreement between the results from both approaches, with better agreement obtained for the low-beam speed regime. 

The organization of the paper is as follows: In section \ref{sec2} a short overview of the generalized weak turbulence theory is presented, with emphasis on the interpretation of the physical origin of different terms in wave-particle kinetic equations. In section \ref{sec3} a concise description of the PIC code is made. Then, in section \ref{sec4} results from both approaches are presented and the comparison between them is carried out. Finally, in section \ref{sec5} we summarize the major findings and present our conclusions.

\section{Weak Turbulence Theory}\label{sec2}

Essential theoretical developments of the generalized weak turbulence theory relevant to the study of plasma emission can be found in the papers by \citet{Yoon06/02, Yoon+12/10, Ziebell+15/06}\@. The theory describes weakly turbulent nonlinear interactions of Langmuir ($L$), ion-sound ($S$), and transverse electromagnetic ($T$) waves, as well as the electrons and protons. The longitudinal electric field wave energy density is given in spectral form by the sum of Langmuir and ion-sound mode intensities, and is defined by
\begin{equation}
\left<\delta E_\parallel^2\right>_{{\bf k},\omega}
=\sum_{\sigma=\pm1}\sum_{\alpha=L,S}I_\alpha^\sigma({\bf k})
\delta(\omega-\sigma\omega_{\bf k}^\alpha)
\label{1}
\end{equation}
where $I_\alpha^\sigma({\bf k})$ represents the individual mode intensity, $\sigma=\pm1$ represents the forward/backward propagation direction of wave modes defined with respect to the electron beam propagation direction, and $\omega_{\bf k}^\alpha$ denotes the wave dispersion relation. The electric and magnetic field wave energy densities for the transverse mode $T$ is expressed in terms of the $T$ mode wave intensity $I_T^\sigma({\bf k})$ by
\begin{equation}
\left<\delta E_\perp^2\right>_{{\bf k}\omega}
=\sum_{\sigma=\pm1}I_T^\sigma({\bf k})
\delta(\omega-\sigma\omega_{\bf k}^T),\qquad
\left<\delta B^2\right>_{{\bf k}\omega}
=\sum_{\sigma=\pm1}\left|\frac{ck}{\omega_{\bf k}^T}\right|^2
I_T^\sigma({\bf k})\delta(\omega-\sigma\omega_{\bf k}^T).
\label{2}
\end{equation}
In the above $\sigma=\pm1$ represents the directions of the wave phase speed with respect to the initial electron beam propagation direction, $\sigma=+1$ representing the forward propagation, while $\sigma=-1$ denote the backward direction. The linear dispersion relations for $L$, $S$, and $T$ modes are given, respectively, by
\begin{equation}
\omega_{\bf k}^L=\omega_{pe}
\left(1+\frac{3}{2}k^2\lambda_D^2\right),\qquad
\omega_{\bf k}^S=\frac{kc_S\left(1+3T_e/T_i\right)^{1/2}}
{(1+k^2\lambda_D^2)^{1/2}},\qquad
\omega_{\bf k}^T=(\omega_{pe}^2+c^2k^2)^{1/2},
\label{3}
\end{equation}
where
\begin{equation}
\omega_{pe}=\left(\frac{4\pi n_ee^2}{m_e}\right)^{1/2},\qquad
\lambda_D=\left(\frac{T_e}{4\pi n_ee^2}\right)^{1/2}
=\frac{v_{th}}{\sqrt{2}\omega_{pe}},\qquad
v_{th}=\left(\frac{2T_e}{m_e}\right)^{1/2}\qquad
c_S=\left(\frac{T_e}{m_i}\right)^{1/2}.
\label{4}
\end{equation}
In the above $\omega_{pe}$ is the plasma frequency, $n_e$, $e$, and $m_e$ being the electron number density, unit electric charge, and electron mass, respectively; $\lambda_D$ represents the Debye length, $T_e$ and $T_i$ being the electron and ion temperatures, respectively; and $v_{th}$ and $c_S$ stand for electron thermal speed and ion-sound speeds, respectively, $m_i$ being the proton mass.

The equations of weak turbulence theory are made of kinetic equations that govern the dynamical evolution of the electron velocity distribution function, $F_e({\bf v},t)$, and spectral wave intensities, $I_L^{\pm}({\bf k},t)$, $I_S^{\pm}({\bf k},t)$, and $I_T^{\pm}({\bf k},t)$,
\begin{equation}
\frac{\partial F_e({\bf v})}{\partial t}=
\frac{\pi e^2}{m_e^2}\sum_{\sigma=\pm1}\int d{\bf k}\,
\frac{\bf k}{k}\cdot\frac{\partial}{\partial{\bf v}}
\,\delta(\sigma\omega_{\bf k}^L-{\bf k}\cdot{\bf v})
\left(\frac{m_e\sigma\omega_{\bf k}^L}{4\pi^2k}\,F_e({\bf v})
+I_L^\sigma({\bf k})\,\frac{\bf k}{k}\cdot
\frac{\partial F_e({\bf v})}{\partial{\bf v}}\right),
\label{5}
\end{equation}
\begin{eqnarray}
\frac{\partial I_L^\sigma({\bf k})}{\partial t}
&=& \frac{4\pi e^2}{m_ek^2}\int d{\bf v}\,
\delta(\sigma\omega_{\bf k}^L-{\bf k}\cdot{\bf v})
\left(n_ee^2F_e({\bf v})+\pi\sigma\omega_{\bf k}^L{\bf k}
\cdot\frac{\partial F_e({\bf v})}{\partial{\bf v}}
\,I_L^\sigma({\bf k})\right)
\nonumber\\
&& +\frac{\pi e^2\sigma\omega_{\bf k}^L}{2T_e^2}
\sum_{\sigma',\sigma''=\pm1}\int d{\bf k}'\,
\frac{\mu_{{\bf k}-{\bf k}'}({\bf k}\cdot{\bf k}')^2}
{k^2{k'}^2|{\bf k}-{\bf k}'|^2}\left(
\sigma\omega_{\bf k}^L\,I_L^{\sigma'}({\bf k}')\,
\frac{I_S^{\sigma''}({\bf k}-{\bf k}')}{\mu_{{\bf k}-{\bf k}'}}\right.
\nonumber\\
&& \left.-\sigma'\omega_{{\bf k}'}^L\,
\frac{I_S^{\sigma''}({\bf k}-{\bf k}')}{\mu_{{\bf k}-{\bf k}'}}
\,I_L^\sigma({\bf k})-\sigma''\omega_{{\bf k}-{\bf k}'}^L
I_L^{\sigma'}({\bf k}')I_L^\sigma({\bf k})\right)
\delta(\sigma\omega_{\bf k}^L-\sigma'\omega_{{\bf k}'}^L
-\sigma''\omega_{{\bf k}-{\bf k}'}^S)
\nonumber\\
&& +\frac{\sigma\omega_{\bf k}^Le^2}{n_em_e^2\omega_{pe}^2}
\sum_{\sigma'}\int d{\bf k}'\int d{\bf v}
\frac{({\bf k}\cdot{\bf k}')^2}{k^2{k'}^2}
\left\{\frac{n_ee^2}{\omega_{pe}^2}
\left[\sigma\omega_{\bf k}^LI_L^{\sigma'}({\bf k}')
-\sigma'\omega_{{\bf k}'}^LI_L^\sigma({\bf k})\right]
\left[F_e({\bf v})+F_i({\bf v})\right]\right.
\nonumber\\
&& \left.+\frac{\pi m_e}{m_i}I_L^{\sigma'}({\bf k}')
I_L^\sigma({\bf k})({\bf k}-{\bf k}')\cdot
\frac{\partial F_i({\bf v})}{\partial{\bf v}}\right\}
\delta[\sigma\omega_{\bf k}^L-\sigma'\omega_{{\bf k}'}^L
-({\bf k}-{\bf k}')\cdot{\bf v}],
\label{6}
\end{eqnarray}
\begin{eqnarray}
\frac{\partial}{\partial t}\frac{I_S^\sigma({\bf k})}
{\mu_{\bf k}} &=& \frac{4\pi\mu_{\bf k}e^2}{m_ek^2}
\int d{\bf v}\,\delta(\sigma\omega_{\bf k}^S-{\bf k}\cdot{\bf v})
\left[n_ee^2\left[F_e({\bf v})+F_i({\bf v})\right]\right.
\nonumber\\
&& \left.+\pi\sigma\omega_{\bf k}^L\left({\bf k}\cdot
\frac{\partial F_e({\bf v})}{\partial{\bf v}}
+\frac{m_e}{m_i}\,{\bf k}\cdot
\frac{\partial F_i({\bf v})}{\partial{\bf v}}\right)
\frac{I_S^\sigma({\bf k})}{\mu_{\bf k}}\right]
\nonumber\\
&& +\frac{\pi e^2\sigma\omega_{\bf k}^L}{4T_e^2}
\sum_{\sigma',\sigma''}\int d{\bf k}'\,
\frac{\mu_{\bf k}\left[{\bf k}'\cdot({\bf k}-{\bf k}')\right]^2}
{k^2{k'}^2|{\bf k}-{\bf k}'|^2}\left(
\sigma\omega_{\bf k}^LI_L^{\sigma'}({\bf k}')
I_L^{\sigma''}({\bf k}-{\bf k}')\right.
\nonumber\\
&& \left.-\sigma'\omega_{{\bf k}'}^L\,I_L^{\sigma''}({\bf k}-{\bf k}')
\,\frac{I_S^\sigma({\bf k})}{\mu_{\bf k}}
-\sigma''\omega_{{\bf k}-{\bf k}'}^L\,I_L^{\sigma'}({\bf k}')
\,\frac{I_S^\sigma({\bf k})}{\mu_{\bf k}}\right)
\delta(\sigma\omega_{\bf k}^S-\sigma'\omega_{{\bf k}'}^L
-\sigma''\omega_{{\bf k}-{\bf k}'}^L),
\label{7}
\end{eqnarray}
\begin{eqnarray}
\frac{\partial}{\partial t}\frac{I_T^\sigma({\bf k})}{2}
&=& \frac{\pi e^2\sigma\omega_{\bf k}^T}{32m_e^2\omega_{pe}^2}
\sum_{\sigma',\sigma''}\int d{\bf k}'\frac{({\bf k}\times{\bf k}')^2}
{k^2{k'}^2|{\bf k}-{\bf k}'|^2}\left(
\frac{k'^2}{\sigma'\omega_{{\bf k}'}^L}
-\frac{|{\bf k}-{\bf k}'|^2}{\sigma''
\omega_{{\bf k}-{\bf k}'}^L}\right)^2
\delta(\sigma\omega_{\bf k}^T-\sigma'\omega_{{\bf k}'}^L
-\sigma''\omega_{{\bf k}-{\bf k}'}^L)
\nonumber\\
&& \times\left(\sigma\omega_{\bf k}^TI_L^{\sigma'}({\bf k}')
I_L^{\sigma''L}({\bf k}-{\bf k}')
-\sigma'\omega_{{\bf k}'}^L\,I_L^{\sigma''}({\bf k}-{\bf k}')
\,\frac{I_T^\sigma({\bf k})}{2}-\sigma''\omega_{{\bf k}-{\bf k}'}^L
\,I_L^{\sigma'}({\bf k}')\,\frac{I_T^\sigma({\bf k})}{2}\right)
\nonumber\\
&& +\frac{\pi e^2\sigma\omega_{\bf k}^T}{4T_e^2}
\sum_{\sigma',\sigma''}\int d{\bf k}'\,
\frac{\mu_{{\bf k}-{\bf k}'}({\bf k}\times{\bf k}')^2}
{k^2{k'}^2|{\bf k}-{\bf k}'|^2}\,
\delta(\sigma\omega_{\bf k}^T-\sigma'\omega_{{\bf k}'}^L
-\sigma''\omega_{{\bf k}-{\bf k}'}^S)
\nonumber\\
&& \times\left(\sigma\omega_{\bf k}^T\,I_L^{\sigma'}({\bf k}')
\,\frac{I_S^{\sigma''}({\bf k}-{\bf k}')}{\mu_{{\bf k}-{\bf k}'}}
-\sigma'\omega_{{\bf k}'}^L\,\frac{I_S^{\sigma''}({\bf k}-{\bf k}')}
{\mu_{{\bf k}-{\bf k}'}^S}\frac{I_T^\sigma({\bf k})}{2}
-\sigma''\omega_{{\bf k}-{\bf k}'}^L\,I_L^{\sigma'}({\bf k}')
\,\frac{I_T^\sigma({\bf k})}{2}\right)
\nonumber\\
&& +\frac{\pi e^2\sigma\omega_{\bf k}^T}{4m_e^2}
\sum_{\sigma',\sigma''}\int d{\bf k}'\,
\frac{|{\bf k}-{\bf k}'|^2}{(\omega_{\bf k}^T)^2
(\omega_{{\bf k}'}^T)^2}\left(1
+\frac{({\bf k}\cdot{\bf k}')^2}{k^2{k'}^2}\right)
\delta(\sigma\omega_{\bf k}^T-\sigma'\omega_{{\bf k}'}^T-\sigma''
\omega_{{\bf k}-{\bf k}'}^L)
\nonumber\\
&& \times\left(\sigma\omega_{\bf k}^T\,I_L^{\sigma''}({\bf k}-{\bf k}')
\,\frac{I_T^{\sigma'}({\bf k}')}{2}
-\sigma'\omega_{{\bf k}'}^T\,I_L^{\sigma''}({\bf k}-{\bf k}')
\,\frac{I_T^\sigma({\bf k})}{2}
-\sigma''\omega_{{\bf k}-{\bf k}'}^L\,
\frac{I_T^{\sigma'}({\bf k}')}{2}
\frac{I_T^\sigma({\bf k})}{2}\right)
\nonumber\\
&& +\frac{\sigma\omega_{\bf k}^Te^2}{2\hat{n}m_e^2\omega_{pe}^2}
\sum_{\sigma'}\int d{\bf k}'\int d{\bf v}\,
\frac{({\bf k}\times{\bf k}')^2}{k^2{k'}^2}
\left[\frac{\hat{n}e^2}{\omega_{pe}^2}
\left(\sigma\omega_{\bf k}^T\,I_L^{\sigma'}({\bf k}')
-\sigma'\omega_{{\bf k}'}^L\,
\frac{I_T^\sigma({\bf k})}{2}\right)
\left[F_e({\bf v})+F_i({\bf v})\right]\right.
\nonumber\\
&& \left.+\pi\,\frac{m_e}{m_i}\,I_L^{\sigma'}({\bf k}')
\,\frac{I_T^\sigma({\bf k})}{2}\,({\bf k}-{\bf k}')\cdot
\frac{\partial F_i({\bf v})}{\partial{\bf v}}\right]
\delta\left[\sigma\omega_{\bf k}^T-\sigma'\omega_{{\bf k}'}^L
-({\bf k}-{\bf k}')\cdot{\bf v}\right],
\label{8}
\end{eqnarray}
where
\begin{eqnarray}
\mu_{\bf k} &=& |k|^3\lambda_{De}^3
\left(\frac{m_e}{m_i}\right)^{1/2}
\left(1+\frac{3T_i}{T_e}\right)^{1/2}.
\label{9}
\end{eqnarray}
In \eqref{5}--\eqref{8}, the dependence of each of the quantities, $F_e({\bf v},t)$, $I_L^{\pm}({\bf k},t)$, $I_S^{\pm}({\bf k},t)$, and $I_T^{\pm}({\bf k},t)$, on time $t$ is implicit.

Physical meanings of various terms in (\ref{5}) -- (\ref{8}) have been expounded in the paper by \citet{Ziebell+15/06}, which is briefly recapitulated here. The electron particle kinetic equation (\ref{5}) is given by the Fokker-Planck type of equation where the linear wave-particle resonance between the electrons and Langmuir turbulence is retained. 

The first term on the right-hand side of Langmuir wave kinetic equation (\ref{6}), which is dictated by $\delta(\sigma\omega_{\bf k}^L-{\bf k}\cdot{\bf v})$, designates the spontaneous and induced emissions of $L$ waves;  the second term that contains the overall three-wave resonance condition $\delta(\sigma\omega_{\bf k}^L-\sigma'\omega_{{\bf k}'}^L-\sigma''\omega_{{\bf k}-{\bf k}'}^S)$ describes the decay/coalescence processes involving two $L$ modes and an $S$ mode; the term dictated by nonlinear wave-particle resonance condition $\delta[\sigma\omega_{\bf k}^L-\sigma'\omega_{{\bf k}'}^L -({\bf k}-{\bf k}')\cdot{\bf v}]$, depicts the spontaneous and induced scattering processes involving two Langmuir waves and the distribution of electrons as well as stationary thermal protons. 

The first term on the right-hand side of $S$ mode wave kinetic equation that contains the factor $\delta(\sigma\omega_{\bf k}^S-{\bf k}\cdot{\bf v})$ corresponds to spontaneous and induced emissions of $S$ waves; nonlinear terms with the factor $\delta(\sigma\omega_{\bf k}^S-\sigma'\omega_{{\bf k}'}^L-\sigma''\omega_{{\bf k}-{\bf k}'}^S)$ depicts the decay/coalescence involving an $S$ mode and two $L$ modes. 

For $T$ mode we have ignored the linear wave-particle resonance term since it is impossible for the electrons to linearly interact with the superluminal $T$ mode. The ${\bf k}'$-integral term on the right-hand side of equation (\ref{8}) that has the three-wave resonance condition $\delta(\sigma\omega_{\bf k}^T-\sigma'\omega_{{\bf k}'}^L-\sigma''\omega_{{\bf k}-{\bf k}'}^L)$ describes the coalescence of two $L$ modes into a $T$ mode, hence, this term is responsible for the harmonic emission at twice the plasma frequency, $\omega\sim2\omega_{pe}$. The term associated with the resonance factor $\delta(\sigma\omega_{\bf k}^T-\sigma'\omega_{{\bf k}'}^L-\sigma''\omega_{{\bf k}-{\bf k}'}^S)$ represents the decay of $L$ mode into an $S$ mode and a $T$ mode at the fundamental plasma frequency, $\omega\sim\omega_{pe}$, which is one of the processes that leads to the fundamental emission. The ${\bf k}'$-integral with the factor $\delta(\sigma\omega_{\bf k}^T-\sigma'\omega_{{\bf k}'}^T-\sigma''\omega_{{\bf k}-{\bf k}'}^L)$ describes the merging of a $T$ mode and an $L$ mode into the next higher harmonic $T$ mode. This is known as the incoherent Raman scattering, and is responsible for higher-harmonic plasma emission, including the third harmonic emission, $\omega\sim3\omega_{pe}$. The term with the condition $\delta[\sigma\omega_{\bf k}^T-\sigma'\omega_{{\bf k}'}^L-({\bf k}-{\bf k}')\cdot{\bf v}]$ attached represents the spontaneous and induced scattering processes involving $T$ and $L$ modes as well as the charged particles. This process can be termed the transformation of $L$ mode into the radiation, mediated by the particles, and is another mechanism responsible for the fundamental emission. 

\citet{Ziebell+15/06} numerically solve the set of equations (\ref{5}) -- (\ref{8}), and by carefully turning various terms on or off, they carried out the detailed diagnostics of each process. The purpose of the present analysis is not to repeat the detailed tasks already carried out by \citet{Ziebell+15/06}, but rather, the major aim of the present work is to validate the efficacy of the WT simulation method by testing the numerical solutions against the PIC code simulation, which is more rigorous.

\section{Particle-in-Cell Simulation}\label{sec3}

We have carried out a series of two-dimensional relativistic and electromagnetic particle-in-cell (PIC) simulations. The simulation box size is $L_X=L_Y=64.83 c/\omega_{pe}=1024 v_{th}/\omega_{pe}$, where $c$ is the speed of light, $v_{th}$ is the electron thermal speed and $\omega_{pe}$ is the electron plasma frequency. The number of grids is $N_X=N_Y=1024$ so that the grid size is $\Delta x = \sqrt{2} \lambda_D$\@.
The simulation time step is $\Delta t=\Delta x /2c$, which satisfies the Courant-Friedrichs-Lewy (CFL) condition, where $\Delta t$ is the time step. 

The number of particles is 200 per grid per species, and we used three kinds of species: protons, background electrons, and beam electrons. The proton-to-electron mass ratio is realistic, $m_p/m_e=1836$, where $m_p$ is the proton mass and $m_e$ is the electron mass. The electron thermal speed is $v_{th}^2/c^2=4.0\times10^{-3}$. The electron temperature is 7 times higher than proton temperature, $T_i/T_e=1/7$, where $T_i$ is the ion (proton) temperature and $T_e$ is the electron temperature. 

The beam consists of $0.1\%$ of the total electron content, namely, $n_b/n_0=10^{-3}$, where $n_b$ is the beam number density and $n_0$ is the electron number density. The Maxwellian beam temperature is the same as the background electron temperature, $T_b=T_e$, where $T_b$ is the electron beam temperature. The plasma to cyclotron frequency ratio is $\omega_{pe}/\Omega_{ce}=100$, where $\Omega_{ce}$ is electron cyclotron frequency. The boundary conditions are periodic for both $X$- and $Y$-directions.

The above described parameters are chosen as in the paper by \citet{Ziebell+15/06}\@. We have carried out three simulations where we have varied the average beam drift speed, $V_b/v_{th}$, from 6, to 8, to 10. Again, these choices are the same as those considered by \citet{Ziebell+15/06}.

\section{Numerical Analysis}\label{sec4}

We solved the equations of GWT turbulence theory, that is, we have carried out the WT simulation, and also carried out the PIC code simulation under the same set of initial conditions so that direct comparisons can be made, which is unprecedented. In the WT simulation we have taken the protons as a stationary background with the two-dimensional velocity distribution function given by
\begin{equation}
F_i({\bf v})=\frac{m_i}{2\pi T_i}
\exp\left(-\frac{m_iv^2}{2T_i}\right),
\label{10}
\end{equation}
where $T_i$ is the proton thermal speed. 

In the PIC code, on the other hand, the protons are free to dynamically evolve, as there are no reasons to fix the protons as a neutralizing background. The initial electron velocity distribution function is the same as that adopted in the analysis by \citet{Ziebell+15/06}, namely, the electrons are composed of a backward drifting background Maxwellian component plus a tenuous forward drifting Maxwellian distribution of electron beam, which in two dimensions, is given by
\begin{equation}
F_e({\bf v},0)=\left(1-\frac{n_b}{n_0}\right)
\frac{m_i}{2\pi T_e}\exp\left(-\frac{m_ev_\perp^2}{2T_e}
-\frac{m_e(v_\parallel+V_0)^2}{2T_e}\right)
+\frac{n_b}{n_0}\frac{m_e}{2\pi T_b}
\exp\left(-\frac{m_ev_\perp^2}{2T_b}
-\frac{m_e(v_\parallel-V_b)^2}{2T_b}\right).
\label{11}
\end{equation} 

We may define the thermal speeds associated with each component in their respective drifting frame, $v_{th}=(2T_e/m_e)^{1/2}$ and $v_{tb}=(2T_b/m_e)^{1/2}$. Here, and $V_0$ and $V_b$ are the drift velocities of the background and the forward beam, respectively, where the background drift velocity $V_0$ is assumed in order to preserve the zero current condition in the proton frame,
\begin{equation}
V_0=\frac{n_bV_b}{n_0-n_b}.
\label{12}
\end{equation}

The PIC code simulation also initialized the electron configuration in the same manner as in the analytical initial velocity distribution function. Initial spectral forms for $L$, $S$, and $T$ mode intensities in the WT simulation are chosen as in \citet{Ziebell+15/06},
\begin{eqnarray}
I_L^\sigma({\bf k},0) &=& \frac{T_e}{4\pi^2}
\frac{1}{1+3k^2\lambda_{De}^2},
\nonumber\\
I_S^\sigma({\bf k},0) &=& \frac{T_e}{4\pi^2}\,
k^2\lambda_{De}^2\left(\frac{1+k^2\lambda_{De}^2}
{1+3k^2\lambda_{De}^2}\right)^{1/2}
\frac{\int d{\bf v}\delta(\sigma\omega_{\bf k}^S
-{\bf k}\cdot{\bf v})(F_e+F_i)}{\int d{\bf v}\delta(\sigma
\omega_{\bf k}^S-{\bf k}\cdot{\bf v})[F_e+(T_e/T_i)F_i]},
\nonumber\\
I_T^\sigma({\bf k},0) &=& \frac{T_e}{2\pi^2}
\frac{1}{1+c^2k^2/\omega_{pe}^2}.
\label{13}
\end{eqnarray}
Of course, in the PIC code simulation, the initial noise is present in the system so that one does not need to specify the initial spectral form for the modes.

The initial input parameters are taken to be the same as that considered by \citet{Ziebell+15/06}\@. These are
\begin{equation}
\frac{1}{n_e\lambda_D^3}=5\times10^{-3},\qquad
T_e=T_b,\qquad\frac{T_i}{T_e}=\frac{1}{7},\qquad
\frac{v_{th}^2}{c^2}=\frac{2T_e}{m_ec^2}=4.0\times10^{-3},
\qquad\frac{n_b}{n_0}=10^{-3}.
\label{14}
\end{equation}
Among the above input parameters, the plasma parameter, $1/(n\lambda_D^3)$ is strictly applicable only for the WT simulation. As we already explained in the previous section, the PIC code simulation was designed with the same specification as (\ref{14}), except for the plasma parameter. In the PIC code, the number of particles per cell is loosely connected to the plasma parameter, as recently demonstrated by \citet{LopezYoon18/10}, but there is no precise way to strictly associate the value of plasma parameter to the PIC code design parameters such as the dimensions, the number of particles per cell, etc. 

We adopt the normalized wave vector, velocity, and time,
\begin{equation}
\frac{{\bf k}v_{th}}{\omega_{pe}},\qquad\frac{\bf v}{v_{th}},\qquad
\omega_{pe}t,
\label{15}
\end{equation}
in plotting the numerical results. Of course, the proton-to-electron mass ratio is realistic, $m_i/m_e=1836$\@. \citet{Ziebell+15/06} considered three cases of normalized forward beam speed, $V_b/v_{th}=6$, 8 and 10. The PIC code simulations are carried out with the same conditions so that direct comparisons can be made. 

In the following, we present the quantitative comparisons of the results obtained by solving the equations of WT theory, that is, WT simulation, and PIC code simulation. In the WT analysis, where we revisited the earlier approach, we have considered a wider velocity space compared to that of \citet{Ziebell+15/06}, with the objective of minimizing the effect of boundary conditions. We present the results in the same format as in the paper by \citet{Ziebell+15/06}, except that we choose to present only the snapshots of electron velocity distribution function, Langmuir wave spectrum, and transverse wave spectrum, or equivalently, the radiation pattern, as these quantities are dynamically important. Ion acoustic mode is excited in both PIC code simulation and WT theory, but as their wave intensities are generally low, we choose not to plot the results. In the PIC code simulation, in particular, fluctuations in the ion acoustic wave frequency range is too weak to distinguish from numerical noise. It should also be noted that the theoretical ion acoustic mode turbulence intensity discussed in the paper by \citet{Ziebell+15/06} is also very low. We thus focus on those quantities that may be meaningfully compared, namely, electron velocity distribution function, high-frequency longitudinal electric field intensity in the Langmuir wave frequency range, and perturbed magnetic field spectrum constructed from PIC simulation, and the corresponding theoretical electron velocity distribution, Langmuir wave intensity, and transverse radiation intensity computed on the basis of WT theory. 

\subsection{Case 1: $V_b/v_{th} = 6$}

\begin{figure}
\centering
\includegraphics[width=0.75\textwidth]{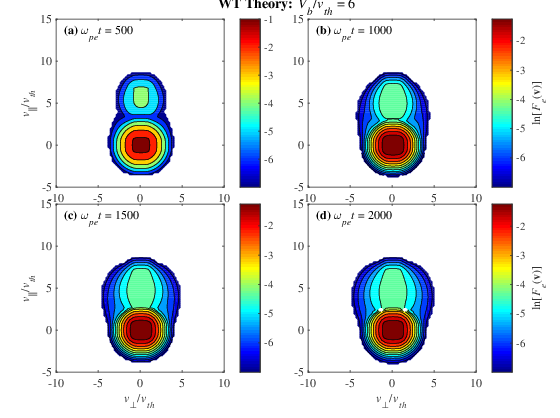}
\includegraphics[width=0.75\textwidth]{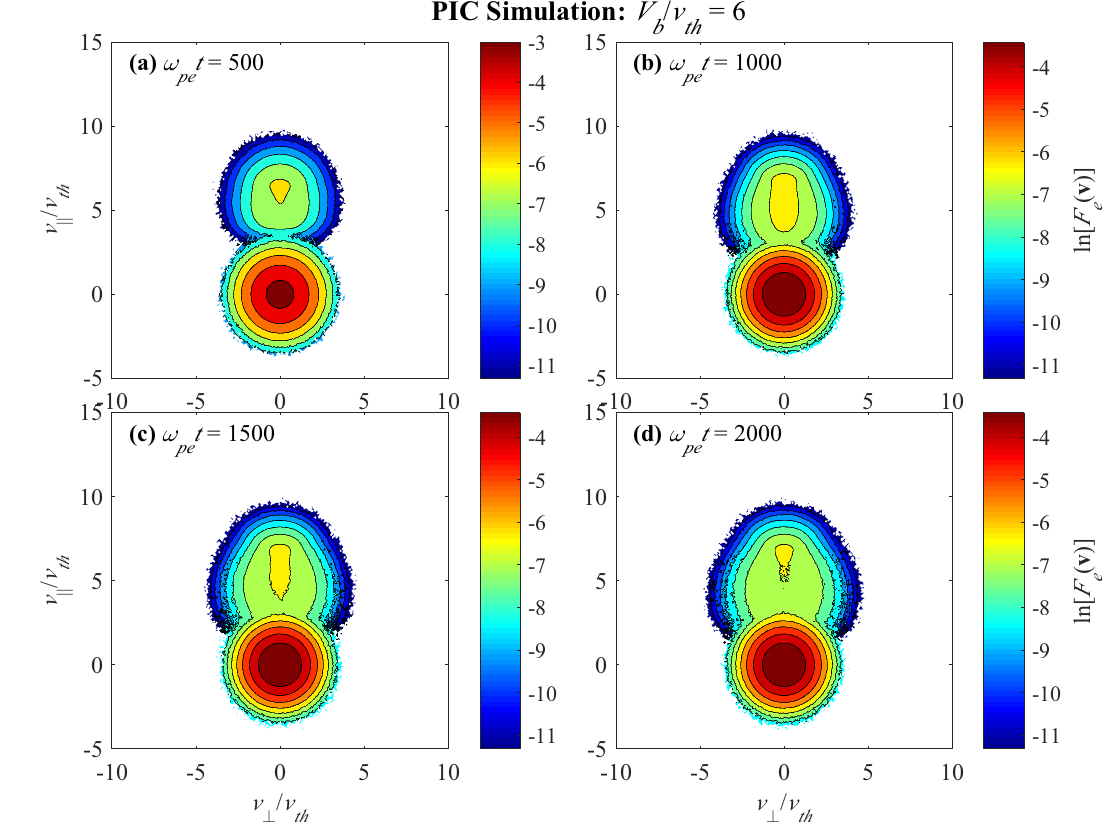}
\caption{Case 1 ($V_b/v_{th}=6$): Time evolution of the electron velocity distribution function (VDF) $F_e({\bf v})$ versus $v_\perp/v_{th}$ and $v_\parallel/v_{th}$, for four different time steps corresponding to $\omega_{pe}t=500$, 1000, 1500, and 2000. Top four panels correspond to WT simulation, while the bottom four panels show results from PIC code simulation.}
\label{F1}
\end{figure}

In Figure \ref{F1} we show the electron velocity distribution function (VDF) computed on the basis of weak turbulence (WT) simulation (top four panels) and the electron velocity phase space distribution constructed from the PIC code simulation (bottom four panels), versus two dimensional normalized perpendicular and parallel speeds, $v_\perp/v_{th}$ and $v_\parallel/v_{th}$, for the first case of $V_b/v_{th}=6$ (case 1). Snapshots of the electron VDF at four different times corresponding to $\omega_{pe}t=500$, 1000, 1500, and 2000, are shown. 

On the basis of the direct comparisons between the theory, or WT simulation, and PIC simulation, it can be seen that there exists an overall quantitative, and even qualitative agreement between the two methods. Specifically, for $\omega_{pe}t=500$, both methods produce similar results in that the beam has undergone a partial plateau formation, but significant positive gradient, $\partial F_e/\partial v_\parallel>0$ still exists along the parallel velocity. As time progresses the positive slope is gradually reduced, until at $\omega_{pe}t=2000$, the plateau formation is almost complete in the case of WT simulation. By contrast, for the PIC code case, a small but finite positive gradient still persists at $\omega_{pe}t=2000$. Despite such a rather insignificant difference, the two methods agree quite well.

\begin{figure}
\centering
\includegraphics[width=0.75\textwidth]{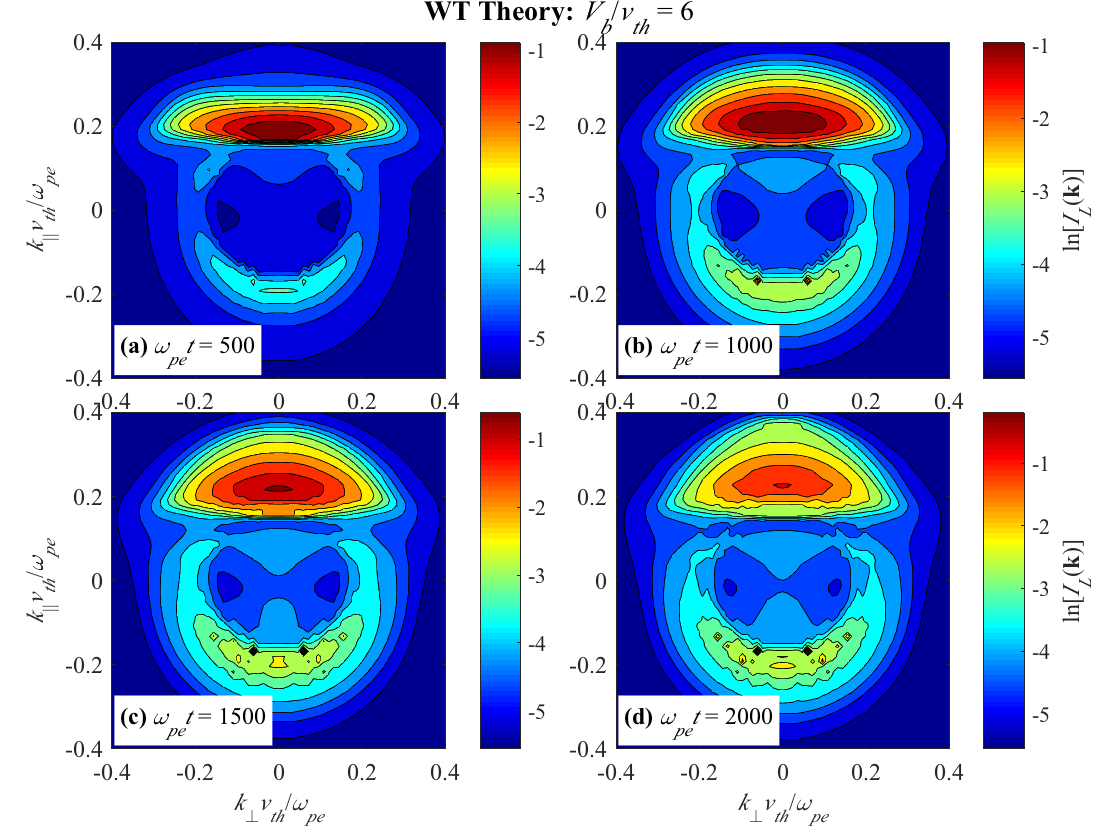}
\includegraphics[width=0.75\textwidth]{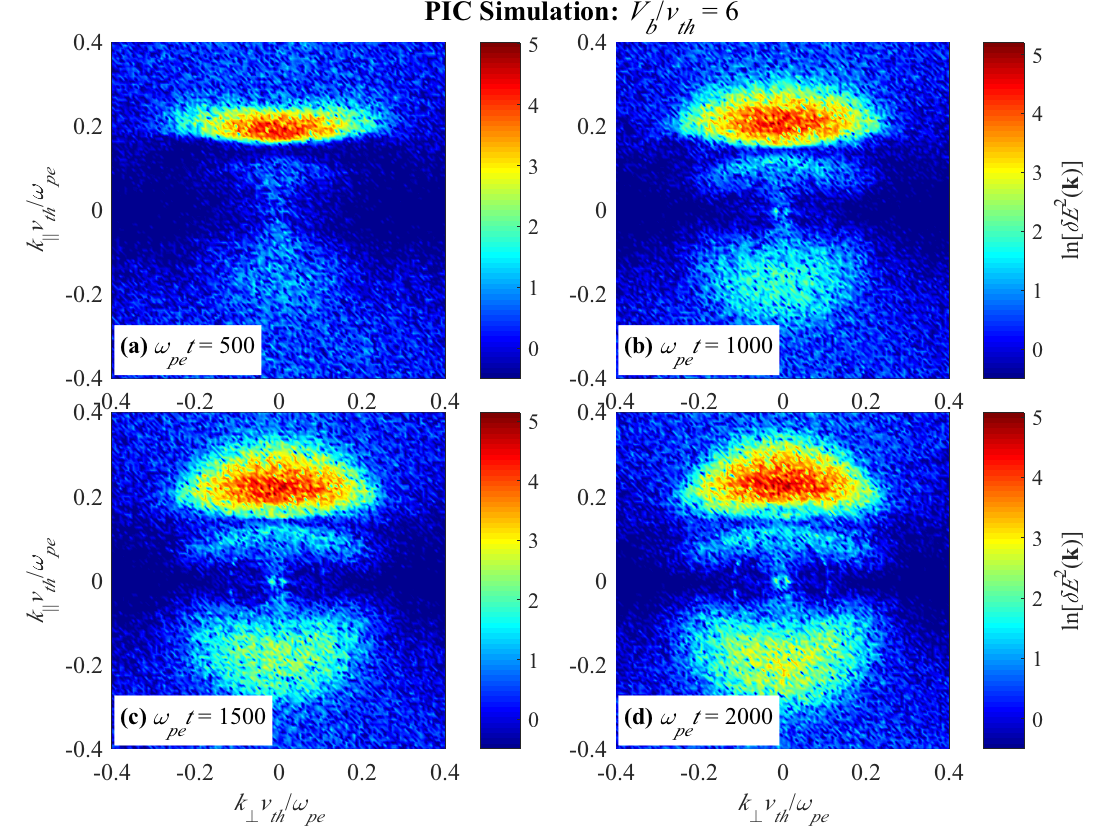}
\caption{Case 1 ($V_b/v_{th}=6$): Time evolution of the Langmuir wave spectral intensity $I_L({\bf k})$, in the case of WT simulation -- top four panels, and total electric field intensity $\delta E^2({\bf k})$, in the case of PIC code simulation -- bottom four panels, versus $k_xv_{th}/\omega_{pe}$ and $k_zv_{th}/\omega_{pe}$, for $\omega_{pe}t=500$, 1000, 1500, and 2000.}
\label{F2}
\end{figure}

The Langmuir turbulence spectrum is shown in Figure \ref{F2}. In the top four panels, the Langmuir spectral intensity $I_L({\bf k})$ is plotted versus $k_\perp v_{th}/\omega_{pe}$ and $k_\parallel v_{th}/\omega_{pe}$. In computing for the forward and backward modes, we restricted ourselves to only positive $k_\parallel>0$ space, but when we plot the final results, we combined both signs of $\sigma=\pm1$ in one panel by showing $I_L^{+1}({\bf k})$ in $k_\parallel>0$ space, while showing $I_L^{-1}({\bf k})$ in the negative $k_\parallel$ space. Thus, the portion of ${\bf k}$ space corresponding to $k_\parallel>0$ should be interpreted as $I_L^{+1}({\bf k})$, while the other half space with $k_\parallel<0$ is the backward $L$ mode, $I_L^{-1}({\bf k})$. For the bottom four panels, which correspond to the simulated electric field spectral intensity, it is not so easy to single out only the contribution from the eigenmode, which $L$ mode is. Instead, we have filtered out the low- and high-frequency fluctuations, and focused only on the spectrum that roughly encompasses the plasma frequency. Note, however, that the spectrum in the vicinity of plasma frequency not only contains the Langmuir mode intensity, which is the linear eigenmode of the plasma, but also nonlinear eigenmode -- see, \citet{Rhee+09/01}\@. The simulated electric field spectrum does not distinguish the two. The total electric field may also contain the radiation mode, but in order to eliminate the transverse component as much as possible, we have selected only the electric field component parallel to the beam propagation direction before implementing the fast Fourier transformation of the simulated electric field. These subtle differences notwithstanding, the comparison between the theoretical intensity and simulated spectrum shows a rather striking resemblance to each other.

Specifically, for early time, $\omega_{pe}t=500$, the enhanced forward-propagating component (the primary $L$) can be seen to be excited in both WT and PIC simulations, which is the result of initial gentle bump-on-tail instability. At this relatively early time the weak backward propagating $L$ mode is also evident. For $\omega_{pe}t=1000$ and beyond, the backscattering of primary $L$ mode into the back-scattered $L$ mode, via a combined three-wave decay process and nonlinear scattering off ions becomes increasingly more visible \citep{ZiebellGaelzerYoon01/09, Ziebell+08/08, Ziebell+12/05, Ziebell+14/11, Ziebell+15/06}\@. In an overall sense, the characteristic time scale of wave evolution and the spectral feature at each stage of time evolution, in particular, the formation of semi-ark shape spectra in both WT and PIC code simulations, are quite consistent. One minor difference is that, whereas in the WT calculation the growth of near $k\sim0$ mode, that is, the Langmuir condensation, is not apparent, the PIC code quite readily indicates that the condensation of Langmuir wave energy to long wavelength mode takes place early on. It has been shown that the effect of spontaneous electron scattering contributes to scattering to the region of low wave numbers, excluding the region $k\simeq0$ \citep{Ziebell+12/05}. The present calculation does include such an effect, but apparently, it is not sufficient to account for the formation of Langmuir condensation effect.

\begin{figure}
\centering
\includegraphics[width=0.75\textwidth]{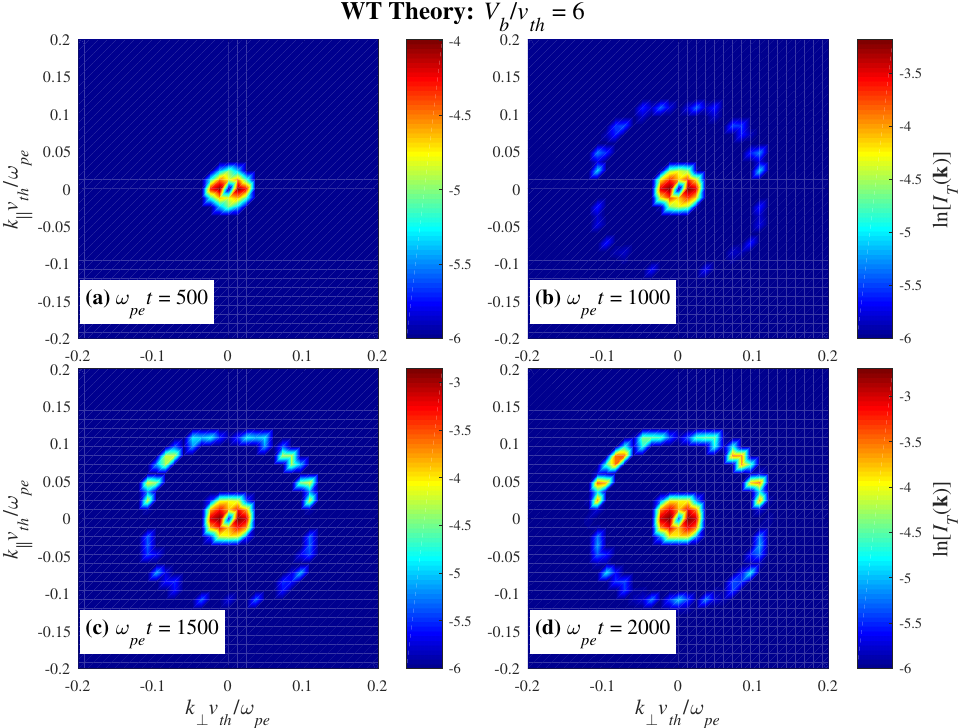}
\includegraphics[width=0.75\textwidth]{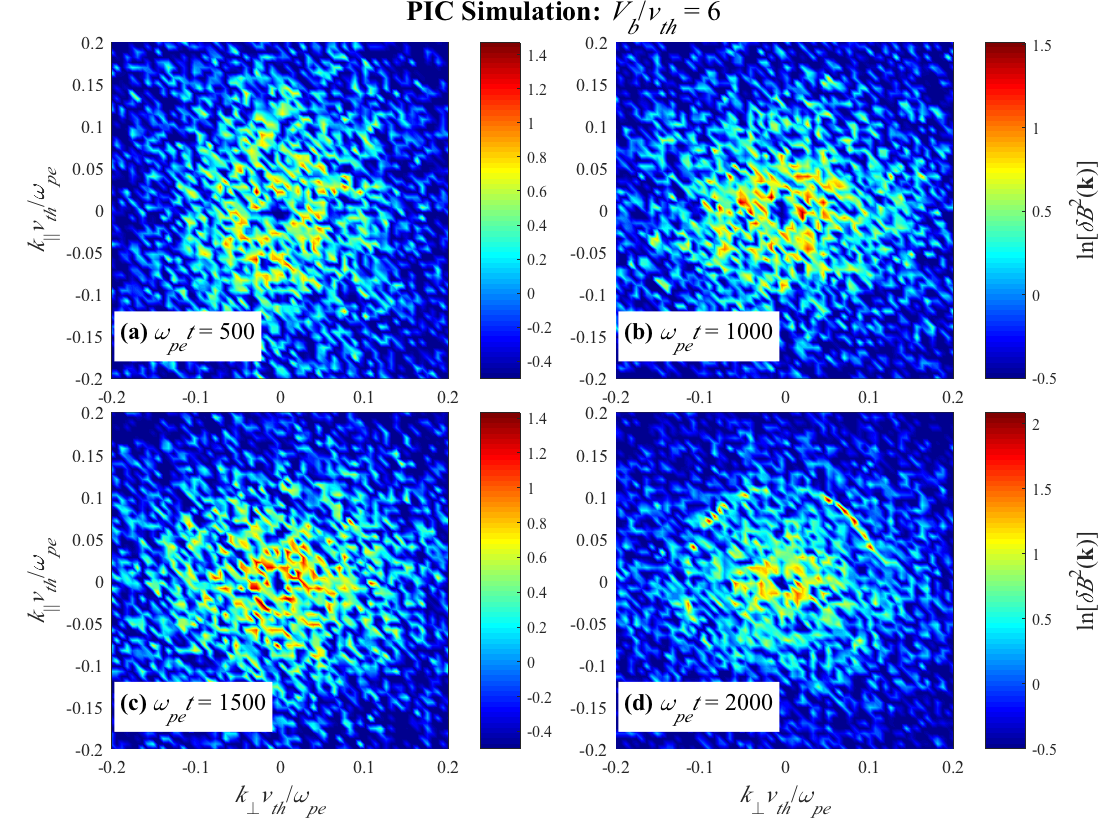}
\caption{Case 1 ($V_b/v_{th}=6$): Time evolution of the radiation spectral intensity $I_T({\bf k})$, in the case of WT simulation -- top four panels, and total magnetic field intensity $\delta B^2({\bf k})$, in the case of PIC code simulation -- bottom four panels, versus $k_xv_{th}/\omega_{pe}$ and $k_zv_{th}/\omega_{pe}$, for $\omega_{pe}t=500$, 1000, 1500, and 2000.}
\label{F3}
\end{figure}

Figure \ref{F3} plots the transverse EM radiation ($T$ mode) spectrum computed on the basis of WT theoretical method (top four panels), and the magnetic field spectrum constructed from PIC simulation (bottom four panels). According to WT calculation, for early time $\omega_{pe}t=500$, finite level of fundamental emission is first generated. Then at $\omega_{pe}t=1000$ and beyond, second harmonic mode gradually appears until at $\omega_{pe}t=2000$, the fundamental/harmonic bi-model radiation pattern is established. In contrast, in the present case of weak beam speed, $V_b/v_{th} = 6$, the magnetic field spectrum is dominated by the noise almost throughout the entire simulated time domain. Only at the very end, $\omega_{pe}t=2000$, does a weak signature of fundamental/harmonic pair emission structure becomes barely discernible. This seems to indicate the need for a higher number of particles per cell in order to reduce the numerical noise.

\subsection{Case 2: $V_b/v_{th} = 8$}

\begin{figure}
\centering
\includegraphics[width=0.75\textwidth]{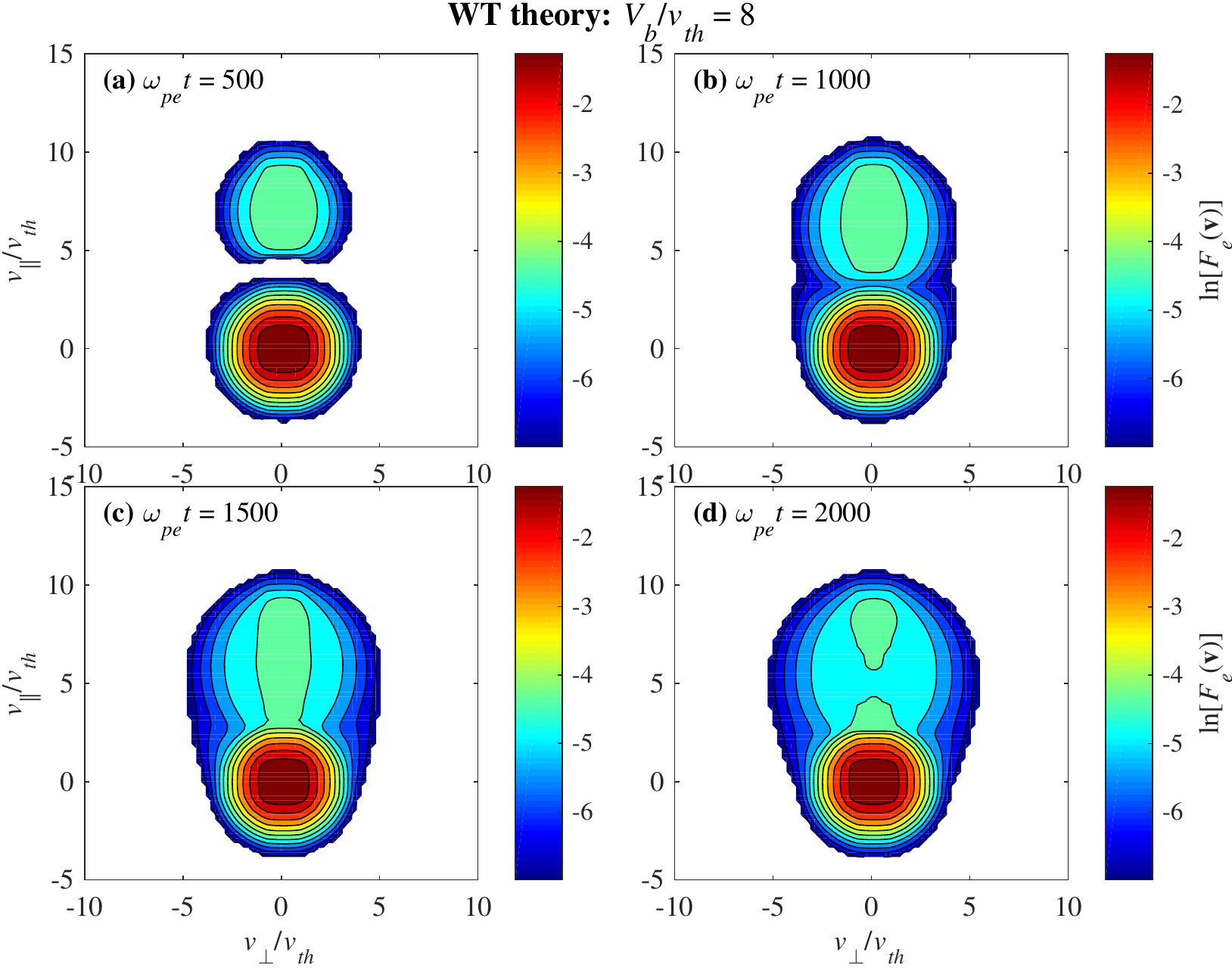}
\includegraphics[width=0.75\textwidth]{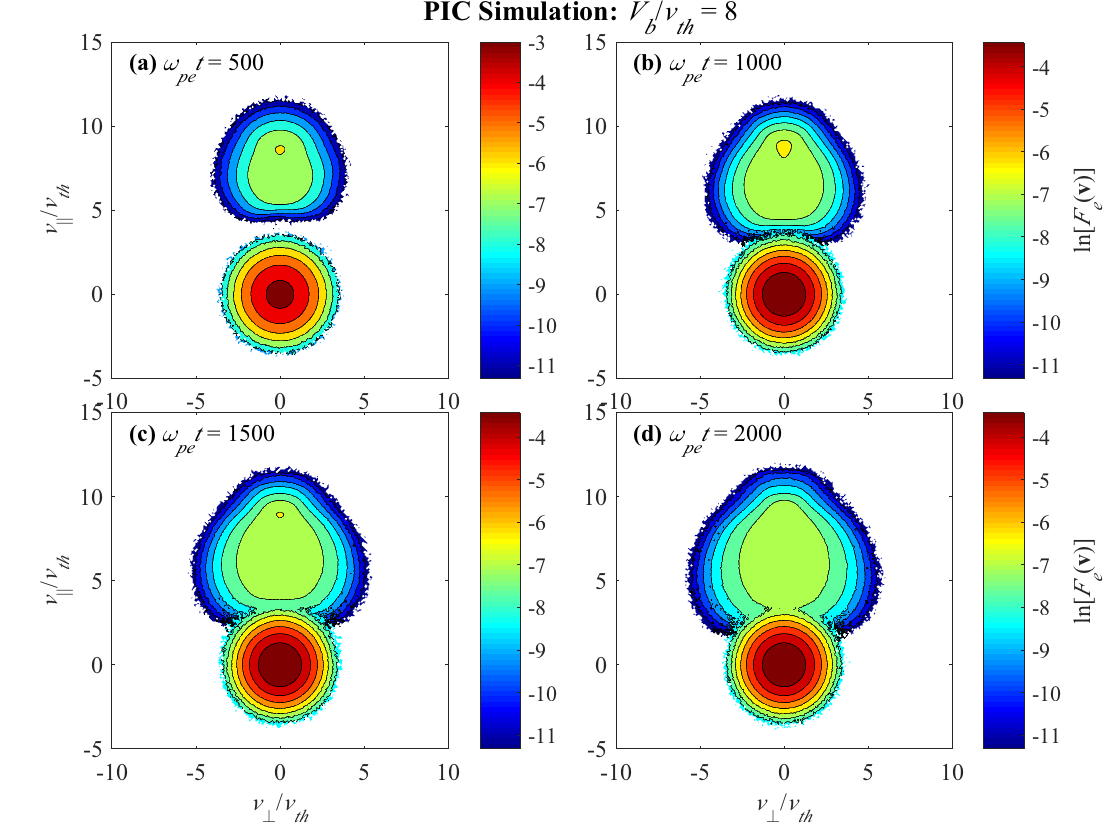}
\caption{Case 2 ($V_b/v_{th}=8$): Electron velocity distribution function (VDF) $F_e({\bf v})$ versus $v_\perp/v_{th}$ and $v_\parallel/v_{th}$, for four different time steps corresponding to $\omega_{pe}t=500$, 1000, 1500, and 2000. Top four panels correspond to WT simulation, while the bottom four panels show results from PIC code simulation.}
\label{F4}
\end{figure}

\citet{Ziebell+15/06} also considered the second case corresponding to $V_b/v_{th} = 8$ (case 2), which we consider next. Figure \ref{F4} displays the electron VDF in the same format as Figure \ref{F1}. As with case 1, the time
evolution of electron VDF is qualitatively similar for both WT and PIC simulations, especially for relatively early time, $\omega_{pe}t=500$. Note, however, that as the system evolves in time, while the qualitative agreement is still maintained, upon close inspection, some quantitative discrepancies become noticeable. Most appreciable is the fact that while the two dimensional structure associated with the theoretically computed electron VDF is defined by elliptical outer contours for the beam population, the PIC simulated electron VDF exhibits a distinct broadening of the beam along $v_\perp$, while the highest parallel velocity portion of the beam population does not suffer from such broadening. Another feature that is noteworthy is that for $\omega_{pe}t=2000$, the theoretical VDF still shows a mild positive parallel derivative associated with the beam, whereas for the simulated VDF such a feature is almost completely gone. In spite of these, it is quite fair to say that the WT simulation features a good overall comparison against PIC simulation.

\begin{figure}
\centering
\includegraphics[width=0.75\textwidth]{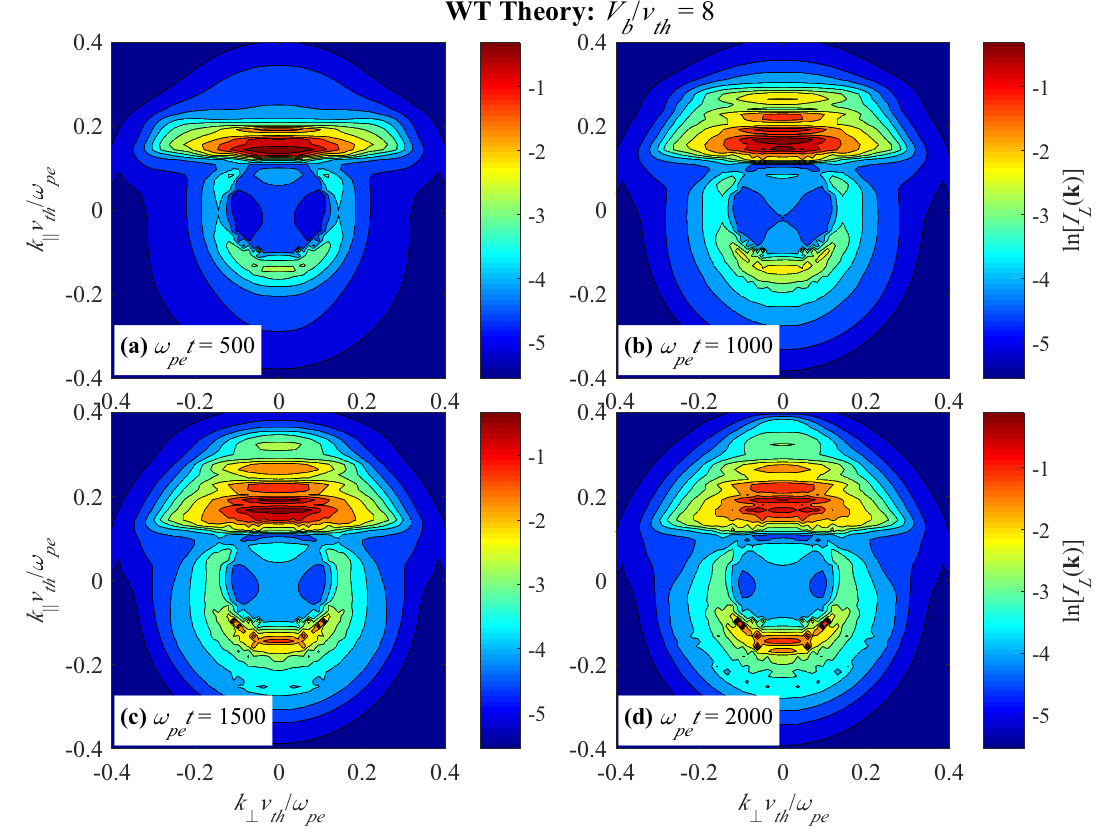}
\includegraphics[width=0.75\textwidth]{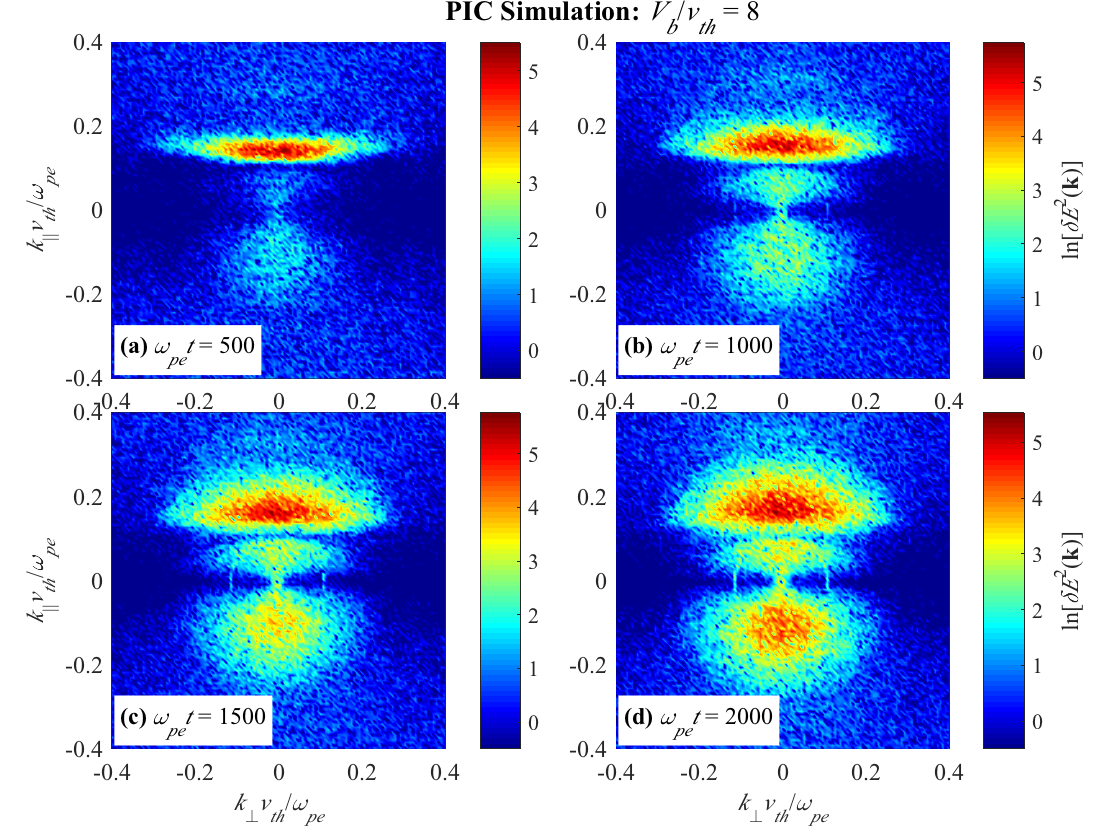}
\caption{Case 2 ($V_b/v_{th}=8$): Langmuir wave spectral intensity $I_L({\bf k})$, in the case of WT simulation -- top four panels, and total electric field intensity $\delta E^2({\bf k})$, in the case of PIC code simulation -- bottom four panels, versus $k_xv_{th}/\omega_{pe}$ and $k_zv_{th}/\omega_{pe}$, for $\omega_{pe}t=500$, 1000, 1500, and 2000.}
\label{F5}
\end{figure}

The Langmuir turbulence spectrum as well as the simulated longitudinal electric field spectrum are plotted in Figure \ref{F5}. Again, despite the subtle differences in the two quantities plotted, as already explained (namely, the theoretical quantity is the Langmuir wave intensity while the simulated quantity is longitudinal electric field fluctuation centered around the plasma oscillation frequency, which may contain both the Langmuir mode as well as the nonlinear eigenmode), the overall agreement is rather remarkable. Upon direct comparison, it is seen that the contours for both the theoretical spectrum and the simulated intensity evolve into more or less ring-like morphology in two dimensional ${\bf k}$ space. For the present case of higher beam speed, more free energy is available for the excitation of Langmuir instability. Consequently, the entire instability and ensuing nonlinear processes develop much more rapidly. This results in many back-and-forth decay and scattering processes, which leads to the multiply-peaked structures in the wave spectrum along $k_\parallel$. This is particularly apparent in the WT simulation. In the PIC code simulation, on the other hand, owing to the inherent noise, the clear delineation of multiple-peak structure along $k_\parallel$ is not so evident. However, upon close examination, especially for $\omega_{pe}t=2000$, the forward propagating longitudinal mode does indeed show a faint evidence for the structure along parallel wave number. As with the first case of $V_b/v_{th}=6$, the PIC code simulation exhibits a rather robust Langmuir condensation phenomenon, while the WT simulation shows very weak or no evidence for Langmuir condensation.

\begin{figure}
\centering
\includegraphics[width=0.75\textwidth]{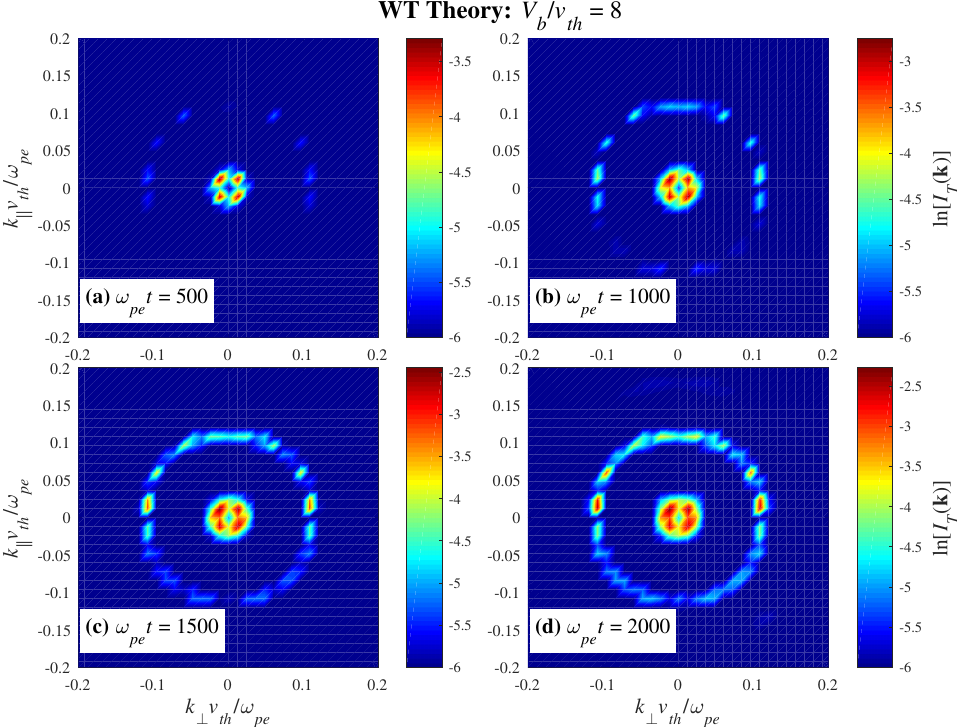}
\includegraphics[width=0.75\textwidth]{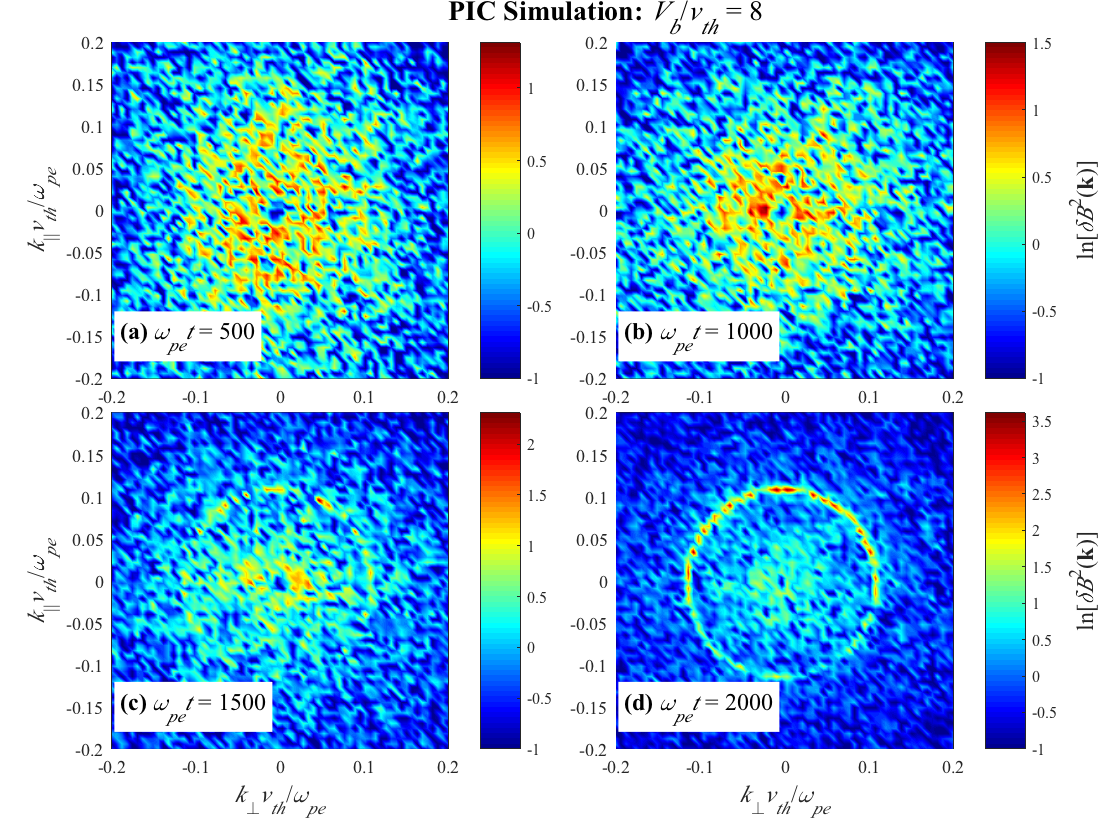}
\caption{Case 2 ($V_b/v_{th}=8$): Radiation spectral intensity $I_T({\bf k})$, in the case of WT simulation -- top four panels, and total magnetic field intensity $\delta B^2({\bf k})$, in the case of PIC code simulation -- bottom four panels, versus $k_xv_{th}/\omega_{pe}$ and $k_zv_{th}/\omega_{pe}$, for $\omega_{pe}t=500$, 1000, 1500, and 2000.}
\label{F6}
\end{figure}

Moving on the transverse EM radiation ($T$ mode) for WT simulation, and the transverse magnetic field fluctuation spectrum for PIC simulation, Figure \ref{F6} plots these spectra. In the present case of $V_b/v_{th}=8$, the fundamental plus weak harmonic emission already takes place at $\omega_{pe}t=500$, in the case of WT calculation. The pair emission pattern gradually and monotonically increases in intensity until the end of the computation, namely, $\omega_{pe}t=2000$. In contrast, as with the previous case, PIC simulation shows no identifiable radiation emission pattern for relatively early times, $\omega_{pe}t=500$ and 1000. Even at $\omega_{pe}t=1500$, the pair emission pattern becomes barely visible with great difficulty. However, at the end of the simulation period, $\omega_{pe}t=2000$, the radiation emission pattern now becomes quite discernible over the background noise. Again, the present PIC code simulation study implies the difficulty in faithfully simulating the plasma emission radiation, and calls for a higher number of particles per cell and thus, quieter simulation, which is, needless to say more computationally demanding, and is beyond the scope of the present work. In this regard, the WT simulation is advantageous, since such an approach is free from the noise issue.

\subsection{Case 3: $V_b/v_{th} = 10$}

\begin{figure}
\centering
\includegraphics[width=0.75\textwidth]{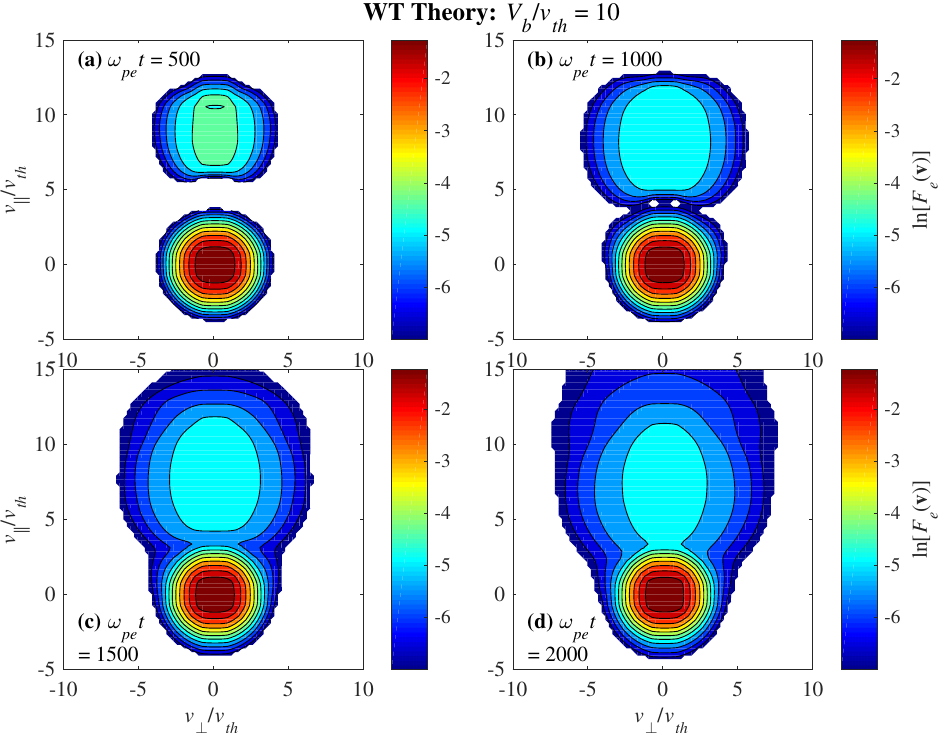}
\includegraphics[width=0.75\textwidth]{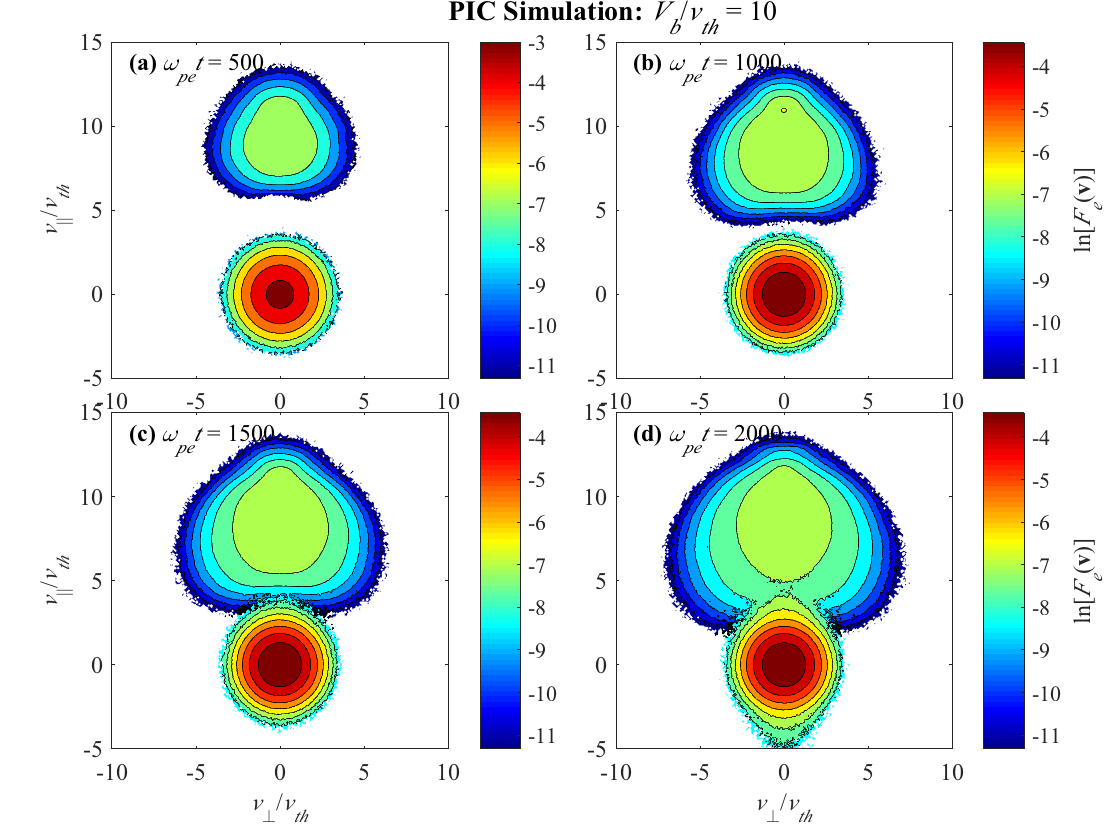}
\caption{Case 3 ($V_b/v_{th}=10$): Electron velocity distribution function (VDF) $F_e({\bf v})$ versus $v_\perp/v_{th}$ and $v_\parallel/v_{th}$, for four different time steps corresponding to $\omega_{pe}t=500$, 1000, 1500, and 2000. Top four panels correspond to WT simulation, while the bottom four panels show results from PIC code simulation.}
\label{F7}
\end{figure}

The third case study is for $V_b/v_{th}=10$ (case 3), which was considered by \citet{Ziebell+15/06}\@. In their paper, the authors speculated that the applicability of weak turbulence theory, and more specifically, the use of weak-growth rate formula inherent in the standard WT formalism might be suspect, but they did not have any standard to verify such a suspicion. For the present relatively high-beam speed, their numerical solution for the electron VDF featured a particularly undesirable aspect of the velocity plateau spreading widely until it reached the boundary, where the boundary effect began to influence the solution. In the present study, we have a benchmark tool to check the validity of WT scheme. In Figure \ref{F7} (top four panels) we plot the electron VDF in the same format as before. We have carefully re-generated the solution with much wider boundary in order to avoid the boundary effects. The case of $\omega_{pe}t=2000$, which appears to show that the beam has spread to the upper boundary of the figure, is actually not a problem since the actual velocity boundary is much wider than what is shown in the figure. In Figure \ref{F7} bottom panels, we display the results of PIC code simulation. As the comparison readily shows, the WT simulation enjoys at least a quantitative agreement, but only in the relative early time periods corresponding to $\omega_{pe}t=500$ and 1000. For $\omega_{pe}t=1500$ and 2000, it is evident that the WT calculation exaggerates the velocity space diffusion of the beam. For PIC code simulation, the beam is spread along $v_\perp$ as in case 2, but the peak velocity portion of the beam along $v_\parallel$ does not evolve much. It is interesting to note that for $\omega_{pe}t=2000$, the PIC simulation indicates parallel acceleration of electrons in both positive and negative portions of $v_\parallel$ axis. This feature is absent in the WT calculation. In an overall sense, while there exist some discrepancies, the qualitative agreement is arguably present, especially for relatively early times. This assessment notwithstanding, the present case 3 study implies that the beam speed of $V_b/v_{th}=10$ may be at the limit of validity for WT theory, at least from the standpoint of electron VDF.

\begin{figure}
\centering
\includegraphics[width=0.75\textwidth]{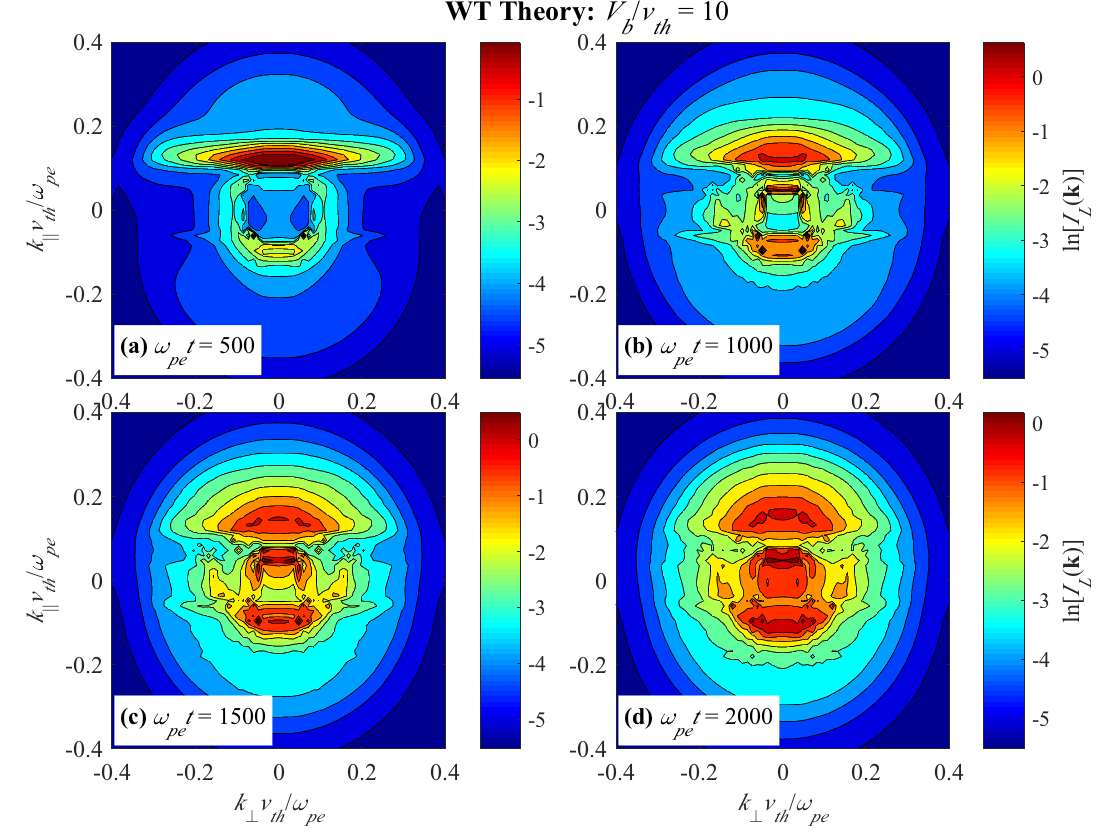}
\includegraphics[width=0.75\textwidth]{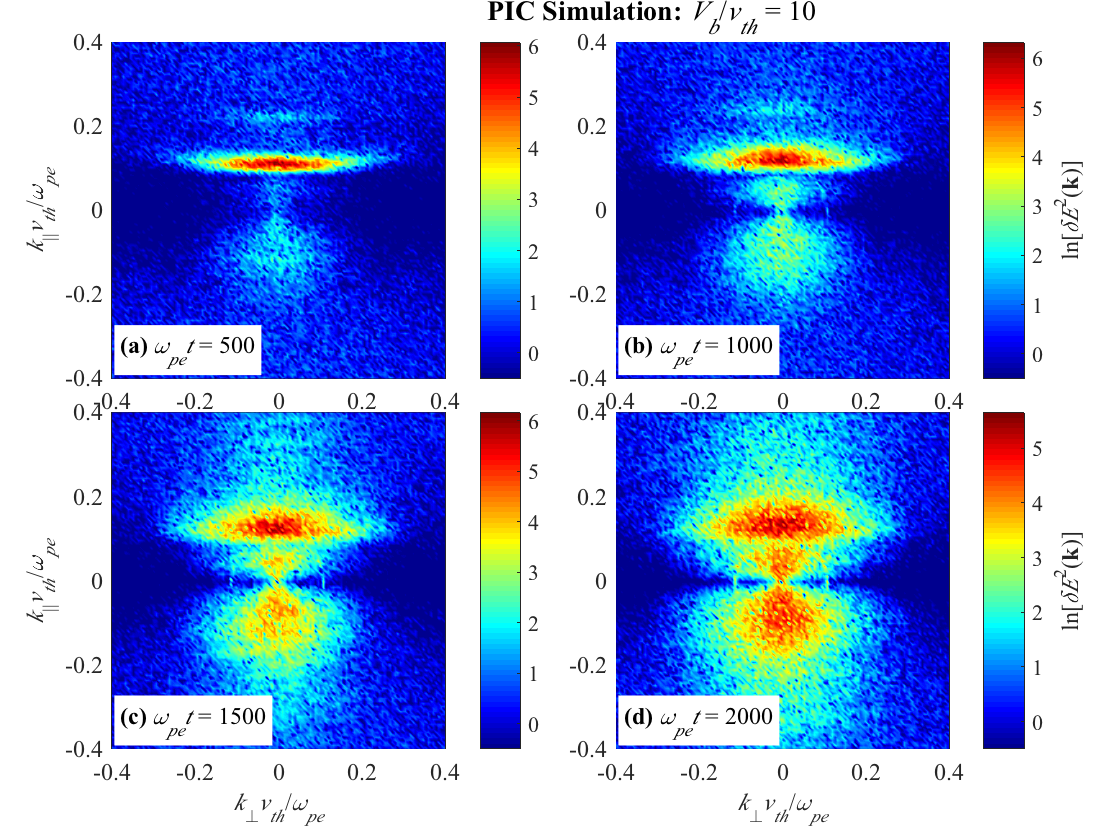}
\caption{Case 3 ($V_b/v_{th}=`0$): Langmuir wave spectral intensity $I_L({\bf k})$, in the case of WT simulation -- top four panels, and total electric field intensity $\delta E^2({\bf k})$, in the case of PIC code simulation -- bottom four panels, versus $k_xv_{th}/\omega_{pe}$ and $k_zv_{th}/\omega_{pe}$, for $\omega_{pe}t=500$, 1000, 1500, and 2000.}
\label{F8}
\end{figure}

However, as for the Langmuir turbulence, the agreement between WT method and PIC code simulation is not that bad. Indeed, as Figure \ref{F8} demonstrates, the Langmuir turbulence spectrum and the simulated longitudinal electric field spectrum are qualitatively similar. The overall morphologies of the two dimensional spectra for both methods somehow produce rather consistent results, including the fact that even the WT simulation generates intense Langmuir condensation, which was missing in the previous two cases.

\begin{figure}
\centering
\includegraphics[width=0.75\textwidth]{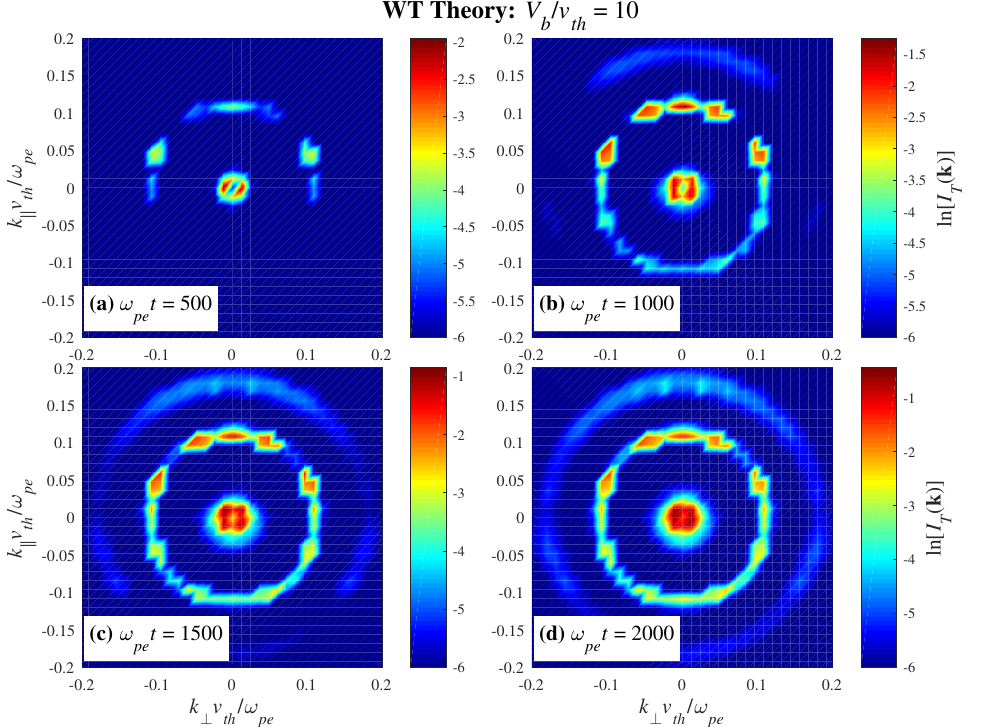}
\includegraphics[width=0.75\textwidth]{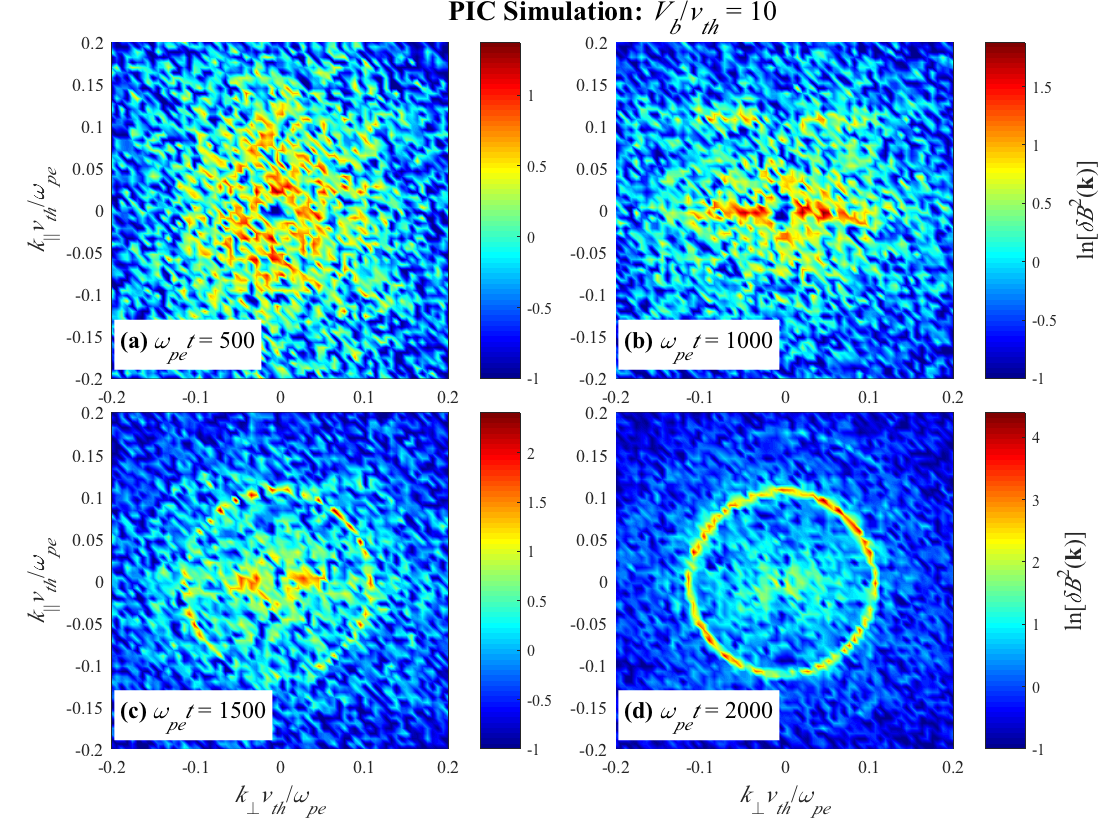}
\caption{Case 3 ($V_b/v_{th}=10$): Radiation spectral intensity $I_T({\bf k})$, in the case of WT simulation -- top four panels, and total magnetic field intensity $\delta B^2({\bf k})$, in the case of PIC code simulation -- bottom four panels, versus $k_xv_{th}/\omega_{pe}$ and $k_zv_{th}/\omega_{pe}$, for $\omega_{pe}t=500$, 1000, 1500, and 2000.}
\label{F9}
\end{figure}

Finally, moving on to the radiation emission pattern, jumping to the final state corresponding to $\omega_{pe}t=2000$, shown in Figure \ref{F9} top and bottom panels, one may immediately appreciate the similarities in the WT versus PIC simulated radiation pattern, which show fundamental/second-harmonic emissions, as well as weak third harmonic emission. Even in the PIC code result, numerical noise notwithstanding, the third harmonic emission is easy to identify. Now, as for relatively early time periods, especially for $\omega_{pe}t=500$ and 1000, the PIC code simulation is still too noisy in order to visually identify and discern clear radiation emission at the harmonics. This contrasts to the WT calculation, which is free of noise problem. By the time the PIC code simulation is carried out to $\omega_{pe}t=1500$, however, the radiation emission pattern begins to manifest itself, albeit, rather faintly.

In the paper by \citet{Ziebell+15/06}, the authors have analyzed the detailed physics of the plasma emission. Specifically, they discussed that the fundamental emission takes place as a result of combined processes of $L$ mode decaying into $T$ and $S$ modes, as well as the scattering involving the
beating of $L$ and $T$ modes mediated by the particles. They mention that both the decay and scattering mechanisms are governed by coupling coefficient of the form,
\begin{equation}
\frac{({\bf k}\times{\bf k}')^2}{k^2{k'}^2}
\propto\sin^2\vartheta,
\label{16}
\end{equation}
where $\vartheta$ represents the angle between the two vectors ${\bf k}'$ and ${\bf k}$. They also argued that the fundamental emission along the direction specified by $\vartheta=0$ should be prohibited, thus resulting in the dipole radiation. In the numerical simulation, whether it be based upon the WT theory, or it is by means of direct PIC code method, the dipolar pattern associated with the fundamental radiation is difficult to discern, since it involves a narrow region around $k\sim0$. 

For the second-harmonic emission, on the other hand, it is well known that the fundamental emission mechanism involves the coalescence of two oppositely traveling Langmuir waves with the coupling coefficient of the form,
\begin{equation}
\frac{({\bf k}\times{\bf k}')^2}{k^2{k'}^2}
\left({k'}^2-|{\bf k}-{\bf k}'|^2\right)^2
\sim\sin^2\vartheta\cos^2\vartheta.
\label{17}
\end{equation}
This implies a quadrupole pattern, but since the radiation emission generally involves multiple wave modes, the above coupling coefficient is to be integrated over ${\bf k}'$, or equivalently, $\vartheta$, hence, the strict quadrupole emission is not evident in reality. 

\citet{Ziebell+15/06} also confirmed the earlier theories of third- and higher harmonic emission \citep{ZheleznyakovZlotnik74/06, Cairns87/10c, Kliem+92/07}\@. The coupling coefficient for the higher-harmonic emission is given by
\begin{equation}
\left(1+\frac{({\bf k}\cdot{\bf k}')^2}{k^2{k'}^2}\right)
\sim1+\cos^2\vartheta.
\label{18}
\end{equation}
\citet{Ziebell+15/06} also reminded the readers that the third-harmonic plasma emission associated with the solar radio bursts is quite rare \citep{TakakuraYousef74/06, Zlotnik+98/03, Brazhenko+12/10}. 

Finally, \citet{Ziebell+15/06} analyzed the details of the various emission mechanisms by artificially turning certain terms in the $T$ wave kinetic equation (\ref{8}) on or off in order to investigate the consequences thereof. By employing such an approach, they confirmed the various theories that have been proposed in the literature concerning the radiation emission mechanisms. Their conclusion still holds, and it is thus unnecessary to repeat their analysis. It should be noted, however, that such a methodology is unique to theoretical approaches such as the WT methodology, since with the PIC code simulation, despite all the rigors inherent to the approach, it is difficult to distinguish underlying individual physical processes, since all are operative simultaneously. This shows that the WT and PIC simulation tools mutually complement each other, and when employed judiciously, may constitute a powerful research tool for the study of solar radio bursts.

\section{Summary and Conclusions}\label{sec5}

The purpose of the present paper has been to investigate the plasma emission process, by making use of two different approaches, and to discuss the compatibility between the results obtained from these two approaches. The plasma emission is generally acknowledged to be the fundamental radiation emission mechanism for solar type II and type III radio bursts phenomena.

In one of the approaches, we have utilized a self-consistent system of coupled equations, obtained using the framework of weak turbulence (WT) theory, in order to study the time evolution of the velocity distribution function of the electrons and of the spectra of electrostatic and transverse waves. The formulation incorporates the effects of different physical mechanisms. In such an approach the roles of different physical processes can be identified unambiguously by turning certain terms on and off, which has in fact been done by \citet{Ziebell+15/06}. The WT formulation, therefore is a convenient tool to test the validity of various theories proposed in the literature for the generation of plasma emission. The complete set of equations can also be solved without {\it a priori} assumptions in order to quantitatively analyze the plasma emission process, as was done by \citet{Ziebell+15/06}. On the other hand, the WT theory, as with any analytical theory, is based upon a series of assumptions, whose limits of applicability have not been clearly defined or tested. 

The second approach employed in the present investigation has been the direct numerical simulation, based upon the particle-in-cell (PIC) paradigm. The PIC simulations rely on a smaller number of theoretical constraints than the approach based on WT theory, as such an approach basically solves the Lorentz equation of motions for a collection of charged particles, plus the Maxwell's equation. Nonetheless, such a numerical approach is not without some shortcomings. For instance, the necessity of discretization and the finite grid size in velocity space and in wave number space, places some limitations on resolution of very small wavelengths. The simulation method also needs to take into account a large number of particles inside a cell in order to reduce the numerical noise, with the consequent burden on the requirements for computational resources. In addition, it is not so easy to make clear diagnostics about the phenomena which occur in the system, since all mechanisms act simultaneously, and cannot be arbitrarily turned on or off for verification of certain physical processes that operate in the system. In contrast, the WT approach naturally lends itself to such manipulations. The WT method requires far less computational resources.

We have made use of these two approaches to study a plasma system containing one ion population and two electron populations, constituted by an initially Maxwellian background and a tenuous beam. We have considered the same parameters, which have been utilized in a previous paper in which the WT equations have been employed in order to discuss in detail the physical mechanisms involved in the plasma emission process \citep{Ziebell+15/06}\@. The WT analysis of the present paper is essentially the same as that of \citet{Ziebell+15/06}, except that we have considered a wider velocity space in order to minimize the boundary effects. The the initial setup for the PIC code simulation is consistent with that of WT analysis by \citet{Ziebell+15/06}\@. Of course, the PIC code assumes additional parameters, such as the grid size and the number of particles per cell, etc., but the physical condition is consistent with that of \citet{Ziebell+15/06}.

The numerical results discussed in detail indicate that the results obtained with the WT approach and with the PIC simulations are largely compatible. Regarding the evolution of the electron distribution function, both approaches show the formation of \textit{plateau} in the beam region, within compatible time scales. The agreement between the WT and the PIC results is more noticeable for the case of lower beam velocity, which is not too surprising, since the WT theory is based upon the assumption of weak wave growth and low wave energy density when compared to the particle thermal energy density. For the intermediate and high beam velocity cases considered in the present paper, we noticed that after the formation of \textit{plateau}, the high velocity part of the beam in the WT results is slightly wider along $v_{\perp}$ direction than in the PIC results. Moreover, the high velocity case (specifically for $V_b/v_{th}=10$) WT calculation resulted in the formation of an extended tail along the forward direction, which is not seen in the PIC simulation results. This appears to be an indication that the high beam velocity case of $V_b/v_{th}=10$, when all other parameters are held constant, corresponds to the limit of applicability of the WT equation.

Regarding the spectra of waves, the WT approach singles out spectra for each eigenmode, which is built into the theory. In contrast, for PIC simulation, the various eigenmodes, such as Langmuir or transverse waves must be carefully interpreted. For instance, in order to contruct the spectrum of Langmuir waves on the basis of PIC simulation, we ave considered only the electric field component along the direction parallel to the beam velocity. This may separate electrostatic waves from transverse waves, but may still retain some amount of intensity associated to nonlinear harmonics of electrostatic waves, hence the interpretation of PIC simulation result, as far as Langmuir waves are concerned, is not without some ambiguities. Nevertheless, the results obtained with the two approaches are largely compatible. Both the WT results and the PIC results show the formation of the primary Langmuir wave peak, with comparable widths along the directions of parallel and perpendicular wave numbers, the growth of backward propagating waves, which are also characterized by consistent widths in wave number space, and also the spread of the primary peak towards the region of smaller wave numbers. A discrepancy remains, mostly in the cases of low and intermediate beam velocities, namely, $V_b/v_{th}=6$ and 8, in that whereas the PIC simulation shows the early appearance of waves for $k\simeq 0$, the so-called Langmuir condensation effect, in the WT results the region of very small values of $k$ is not quite attained, until the final computational time attained in our analysis. In the case of higher beam velocity which has been considered, namely, $V_b/v_{th}=10$, on the other hand, the WT results also show some growth of waves at $k\simeq 0$. The reason for this localized discrepancy between the two approaches is not yet completely understood. In an overall sense, however, the agreement between WT and PIC methods are more consistent than the electron velocity distribution, which is interesting.

In order to obtain information about the spectra of electromagnetic waves in the case of PIC simulation results, we have taken into account the total magnetic field intensity. The spectrum of the magnetic field fluctuations is then used for comparison with the spectrum of transverse waves computed on the basis of WT method. The comparative analysis produced largely favorable results, although because of numerical noise associated with the PIC code, and because of the generally low level of radiation, the relatively early time results do not clearly show easily identifiable plasma emission. The compatibility between WT and PIC results was seen to improve with the increase of beam velocity. Clear demonstration of the radiation emission was difficult to show with the limited numerical setup adopted in the PIC code. This implies that the simulation of plasma emission requires large number of particles per cell in order to reduce the numerical noise, which requires high computational resources.

To conclude, the PIC code simulation is supposed to be more rigorous, but it necessitates computationally intense efforts. In contrast, while the WT theory is a reduced approach, the comparative analysis presented herewith provided evidence that suggests that the use of WT theory can be reliable, if it is carefully applied to a parameter regime for which the theory is valid. The present investigation has focused on three examples, but more systematical statistical survey of the parameter space could be carried out in order to further establish the region of validity of WT approach. The present paper also indicates the possibility of improved agreement between the WT approach and the PIC simulation approach if the numerical noise can be reduced in the PIC simulations. In short, we find that both the WT theory and PIC code simulations are useful research tools in the fundamental study of solar radio bursts problem, as they are mutually complementary.

\acknowledgments

The present research was supported in part by the BK21 plus grant from the National Research Foundation (NRF) of the Republic of Korea, to Kyung Hee University.
L.F.Z. and R.G. acknowledge support provided by Conselho Nacional de Desenvolvimento Cient\'{i}fico e Tecnol\'{o}gico (CNPq), grants No. 304363/2014-6 and 307626/2015-6\@.
This study was also financed in part by the Coordena\c{c}\~{a}o de Aperfei\c{c}oamento de Pessoal de N\'ivel Superior - Brasil (CAPES) - Finance Code 001.
P.H.Y. acknowledges the grant from the GFT Charity, Inc., to the University of Maryland.
This work was partly carried out while P.H.Y. was visiting Korea Astronomy and Space Science Institute (KASI).
E.L. acknowledges support by Space Core Technology Development Program through the NRF of Korea funded by Ministry of Science and ICT (NRF-2017M1A3A3A02016781).

\bibliography{Lee+18_ApJ}

\end{document}